\documentclass[final,12pt]{elsarticle}

\usepackage[utf8]{inputenc}
\usepackage[T1]{fontenc}

\usepackage{graphicx}
\usepackage{amssymb}

\usepackage{rotating}
\usepackage{multirow}
\usepackage{booktabs}

\usepackage{siunitx}

\usepackage{array}
\usepackage{ragged2e}
\newcolumntype{P}[1]{>{\RaggedRight\hspace{0pt}}p{#1}}

\usepackage{tikz}
\usetikzlibrary{arrows,shapes,positioning,shadows,trees}
\tikzset{
  basic/.style  = {draw, text width=2cm, drop shadow, font=\sffamily, rectangle},
  root/.style   = {basic, rounded corners=2pt, thin, align=center,
                   fill=blue!60},
  level 2/.style = {basic, rounded corners=6pt, thin,align=center, fill=blue!60,
                   text width=8em},
  level 3/.style = {basic, thin, align=left, fill=pink!60, text width=6.5em}
}

\usepackage{footnote}
\makesavenoteenv{table}

\usepackage{pgfplotstable}
\usepgfplotslibrary{fillbetween}
\usetikzlibrary{intersections}
\usetikzlibrary{matrix}

\usepackage{comment}

\pgfplotsset{compat=1.14}

\usepackage{enumitem}

\usepackage{colortbl}
\usepackage{etoolbox}
\usepackage{tablefootnote}
\usepackage[hang]{footmisc}
\usepackage[export]{adjustbox}
\usepackage{ulem}

\usepackage[colorlinks,breaklinks]{hyperref}
\hypersetup{
	colorlinks,
	linkcolor   = {blue!50!black},
	citecolor   = {blue!50!black},
	urlcolor    = {blue!80!black},
	pdftitle    = {TBC},
	pdfauthor   = {All authors},
	pdfcreator  = {LaTeX (TeX Live)}
}

\journal{Energy}

\begin{document}

\begin{frontmatter}


\title{Which model features matter? An experimental approach to evaluate power market modeling choices}



\author[label1]{Kais Siala}
\address[label1]{Chair of Renewable and Sustainable Energy Systems, Technical University of Munich}
\cortext[cor1]{Corresponding author: kais.siala@tum.de}

\author[label2]{Mathias Mier}
\address[label2]{ifo Institute for Economic Research at the University of Munich}

\author[label4]{Lukas Schmidt}
\address[label4]{Institute of Energy Economics and Department of Economics, University of Cologne}

\author[label5]{Laura Torralba-Díaz}
\address[label5]{Institute of Energy Economics and Rational Energy Use (IER), University of Stuttgart}

\author[label3]{Siamak Sheykhha}
\address[label3]{Institute for Future Energy Consumer Needs and Behavior (FCN), School of Business and Economics / E.ON Energy Research Center, RWTH Aachen University}

\author[label5]{Georgios Savvidis}

\begin{abstract}
A novel experimental approach of inter- and intramodel comparisons is conducted with five power market models to give recommendations for modelers working on decarbonization pathways of Europe until 2050. The experiments investigate the impact of model type (optimization vs. simulation), planning horizon (intertemporal vs. myopic), temporal resolution (8760 vs. 384 hours), and spatial resolution (28 countries vs. 12 mega-regions).
The model type fundamentally determines the evolution of capacity expansion. Planning horizon (assumed foresight of firms) plays a minor role for scenarios with high carbon prices. For low carbon prices in turn, results from myopic models deviate considerably from those of intertemporal models. Lower temporal and spatial resolutions foster wind power via storage and via neglected transmission boundaries, respectively.
Using simulation instead of optimization frameworks, a shorter planning horizon of firms, or lower temporal and spatial resolutions might be necessary to reduce the computational complexity. This paper delivers recommendations on how to limit the discrepancies in such cases.

\end{abstract}


\end{frontmatter}

\section{Introduction}

The European energy system has grown in complexity over the past decades: Liberalization broke up integrated energy suppliers and led to a variety of energy market participants. Decarbonization efforts continue to transform the electricity supply~\cite{Mier20}, particularly towards a more spatially distributed system through the rise of renewable energy sources. Furthermore, electricity demand is set to increase since electrification is a major means of decarbonizing the transport and heat sectors~\cite{Helgeson2020}. Within this increasingly complex system, numerical power market models gain importance in assessing future developments and the impact of different policies. In order to provide robust, evidence-based policy recommendations, a fundamental understanding of the numerical models is crucial. 


Power market models are increasingly sophisticated and include a variety of features with superposed, often contradictory effects \cite{pfenninger2014}. Additionally, there are multiple model formulations, different data sources, and various scenarios that can be used to answer the same policy question. To gain more clarity on the behavior of models and avoid contradictory recommendations, it is necessary to conduct model comparisons, preferably with harmonized features, same input data, and consistent scenarios. The harmonization of features, data, and scenarios reduces the differences between the models to the core features and renders the causal inference possible, which ultimately increases the confidence in the modeling methods.

Five power market model frameworks are compared using harmonized input data and features. All models are dispatch and investment models that expand generation capacity in the European power market (EU27 plus Norway, Switzerland, and United Kingdom, without Cyprus and Malta) from 2015 (base year) to 2050 (model horizon) in 5-year steps. The harmonization process starts with the handling of a joint database\footnote{The database was developed during the 4NEMO project and is available upon request under mier@ifo.de} to ensure valid results. Several optional features are then deactivated to create frameworks with similar capabilities. The data and feature harmonization is crucial because it allows the attribution of remaining deviations in results to four key features of power markets models: model type, planning horizon, temporal resolution, and spatial resolution. Finally, differences between the models are assessed in several experiments, each focusing on one of those features.

The four above mentioned features are critical for power market models. Model type determines whether firm behavior is reflected by profit maximization or cost minimization (i.e., optimization), or by decision heuristics (i.e., simulation). Planning horizon covers the assumed foresight of firms, and is either myopic (i.e. short-sighted, dealing with each time period separately) or intertemporal (taking into account all time periods simultaneously). Whereas model type and planning horizon reflect general stakeholder behavior, temporal and spatial resolutions have an impact on the level of detail and the computational complexity. Supply from intermittent renewable energy sources and electricity demand fluctuate over space and time. Storage and transmission, which can balance these fluctuations, are limited. As a consequence, power market models need to reflect these constraints by striking a balance between the model resolution and its solving time. 

The main results are threefold. First, the used simulation model with its current formulation is both slow in exiting conventional technologies (e.g. nuclear and coal) and fast in adopting new technologies, even when their usage is not economically profitable yet. Second, myopic models perform well in comparison to intertemporal models. Although intertemporal models tend to wait longer to exploit sites with high variable renewable energy (VRE) potentials using efficient technologies, overall VRE expansion is similar by 2050. Third, wind output is higher for lower temporal and spatial resolutions due to overestimated storage capabilities when using typical time steps (in contrast to 8760 hours of a year) and missing transmission line contingencies within grouped countries. For the sake of transparency, the harmonized input data as well as the derived results are provided. Hence, other modelers are able to benchmark their models against the used ones. 

The next section explores related literature and compares the adopted approach to other model comparisons. Section \ref{sec:method} describes the model frameworks, model features, and the data harmonization process. Section \ref{sec:experiments} describes the experiments with their settings and results. Section \ref{sec:conclusion} concludes.

\section{Literature}

At the core of this paper is a novel experimental approach to compare numerical power market models, which has not been done before to that extent with regard to feature and data harmonization. The five models involved were part of the 4NEMO project\footnote{\textit{Research Network for the Development of New Methods in Energy System Modeling (4NEMO)} funded by the Federal Ministry for Economic Affairs and Energy of Germany (BMWi) under grant number 0324008A.}, which provided the necessary resources to conduct the analysis. Table \ref{tab:LR} gives an overview of reviews, as well as intramodel and intermodel comparisons in the field of energy system modeling. To distinguish between intramodel and intermodel comparisons, the difference between frameworks and models is first clarified: whereas a framework is an abstract set of code, a model is an instance of that code combined with actual data for a specific problem. Hence, intramodel comparisons rely on a single framework with multiple model instances, whereas intermodel comparisons involve different frameworks.

The table also assesses the level of data harmonization, the number of considered model frameworks, and whether or not the papers analyze the impact of model type, planning horizon, temporal resolution, and spatial resolution, as is done in this paper.

Reviews discuss qualitative properties and usually cover a wide spectrum of model frameworks with multiple variations. They support decision-makers in selecting suitable model frameworks for answering specific policy and research questions~\cite{gacitua2018comprehensive}. In general, they classify models~\cite{hall2016review} or identify current challenges and trends in the literature~\cite{connolly2010}. A better overview of the model landscape further aids researchers in identifying research gaps~\cite{savvidis2019gap}. However, reviews cannot quantify the impact of different model, feature, or data choices on the outcomes. To this end, model comparisons are carried out.

\begin{table}[ht!]
    \centering
    \caption{Literature review. Dashes (--) mean that the attributes are not applicable.}
    \label{tab:LR}
    \vspace{0.3cm}
    \footnotesize
    \begin{tabular}{p{0.4cm} p{1.8cm} p{1cm} p{0.6cm} p{1.1cm} p{1.3cm} p{1.7cm} p{1.7cm}}
    
    Ref. & Comparison type & Harmo-nization & Nr.\footnote{Number of models frameworks compared} & Model types\footnote{Abbreviations: E~=~Econometric, O~=~Optimization, S~=~Simulation} & Planning horizon\footnote{Abbreviations: I~=~Intertemporal, M~=~Myopic} & Temporal resolution\footnote{A cross (X) means that the paper discusses the impact of varying the temporal resolution.} & Spatial resolution\footnote{A cross (X) means that the paper discusses the impact of varying the spatial resolution.}\\\toprule
        
    \cite{gacitua2018comprehensive} & Review & -- & 21 & O & I, M & -- & -- \\[2pt]
        
    \cite{hall2016review} & Review & -- & 22 & O, S & I, M & X & X \\[2pt] 
        
    
    \cite{connolly2010} & Review & -- & 37 & O, S & I, M & X & X \\[2pt] 
        
    \cite{savvidis2019gap} & Review & -- & 40 & O, S & I, M & -- & -- \\[2pt]
    \midrule
    
    \cite{ommen2014comparison} & Intramodel & -- & 1 & O & -- & -- & -- \\[2pt] 
        
    \cite{pfenninger2017dealing} & Intramodel & -- & 1 & O & M & X & -- \\[2pt] 
        
    \cite{nahmacher16} & Intramodel & -- & 1 & O & I & X & -- \\[2pt]
        
       
        
    \cite{frew2016temporal} & Intramodel & -- & 1 & O & M & X & X \\[2pt] 

    \cite{priesmann2019complex} & Intramodel & -- & 1 & O & M & X & X \\[2pt]

    \cite{diaz2019importance} & Intramodel & -- & 1 & O & M & X & -- \\[2pt] 
    
    \midrule
    
    \cite{van2009comparison} & Intermodel & Low & 6 & E, O, S  & -- & -- & -- \\[2pt] 
    
    \cite{capros2014european} & Intermodel & Low & 7 & E, O, S & I, M & -- & -- \\[2pt] 
    
    \cite{forster2013european} & Intermodel & Low & 10 & O, S &  I & -- & -- \\[2pt] 
    
    \cite{knopf2013beyond} & Intermodel & Low & 10 & O, S & I & X & X \\[2pt] 

    \cite{deane2012soft} & Intermodel & High & 2 & O & I & X & -- \\[2pt] 
        
    \cite{kiviluoma2018comparison} & Intermodel & High & 2 & O & I & X & -- \\[2pt]

    \cite{poncelet2016impact} &  Intermodel & High & 2 & O & M & X & -- \\[2pt] 
        
        
        
    \cite{gils2019comparison} & Intermodel & High & 4 & O & M & X & X \\[2pt]
    \bottomrule
    \end{tabular}
\end{table}

Intramodel comparisons apply several model configurations to a single model framework and evaluate their effects on the model outputs. \citet{ommen2014comparison} compare different complexity levels of modeling formulation, i.e., linear, mixed integer, and non-linear, within a generic dispatch model framework. \citet{pfenninger2017dealing} and \citet{nahmacher16} evaluate the impact of the selection of representative time slices, while \citet{frew2016temporal} and \citet{priesmann2019complex} assess the trade-off between spatial and temporal resolution systematically. \citet{diaz2019importance} analyze the importance of modeling complexity in long-term energy planning models in terms of time resolution, operational inflexibility and uncertainty regarding fossil fuel prices. A clear advantage of intramodel comparisons is that there is no need for data harmonization. However, considering only one framework neglects the effects of its inherent characteristics.

Intermodel comparisons are carried out on model frameworks and help to evaluate the robustness of findings. Some intermodel comparisons derive robust climate change mitigation trends and strategies that remain valid regardless of the applied model framework, such as \citet{van2009comparison} comparing bottom-up and top-down mitigation potentials estimates, \citet{capros2014european} applying different technological and policy choices, \citet{forster2013european} looking into European energy efficiency, and \citet{knopf2013beyond} looking into decarbonization strategies beyond 2020. Those papers involve a high number of models, but the level of input data harmonization is lower in contrast to the experimental approach presented here. Other intermodel comparisons analyze modeling details by comparing only two (examples: \citet{deane2012soft} about soft-linking, \citet{kiviluoma2018comparison} on the value of renewable generation, and \citet{poncelet2016impact} on the impact of modeling detail) or four models \cite{gils2019comparison} with highly harmonized data. The latter is the most similar approach from literature to this paper. It presents a systematic model comparison between energy system models based on linear optimization with focus on temporal and spatial resolution as well as sector coupling. The analysis includes three scenarios for the German power supply system in 2050. Unlike this approach, the comparison presented here includes a simulation model with heuristic-based investment decisions alongside optimization models with cost-minimizing objective functions. By extending the planning horizon to more than one year, this paper also investigates the impact of the choice between a myopic and an intertemporal approach.

Based on the evidence from the literature, the following conclusions can be drawn. First, most model comparisons are either purely intramodel or intermodel, but not both. Second, while the impact of the temporal resolution is prominent in most papers, the other categories are not well researched. Investigating the spatial resolution features only in a few, and only \citet{capros2014european} evaluate the impact of the planning horizon. Different model types are only compared in intermodel comparisons with low harmonization levels. Third, the level of harmonization for intermodel comparisons of more than two models is typically low, with \citet{gils2019comparison} being the only exception. This paper contributes to the prevalent literature by quantifying the impacts of model type, planning horizon, and temporal and spatial resolutions in a highly harmonized intermodel comparison. Thereby, it fills the described gaps in the existing literature.

\section{Frameworks, data, and model features}
\label{sec:method}
This section first introduces the models involved in the model comparisons. This is followed by a description of the data. Finally, the model features are classified into major features, deactivated features, and remaining differences.

\subsection{Model frameworks}
\label{subsec:frameworks}

All considered frameworks (see Table \ref{tab:frameworks}) are partial equilibrium models that cover (most of) the interconnected European power markets (28 countries: EU27 excluding Cyprus and Malta, including Norway, Switzerland, and the United Kingdom). Trade between markets is possible via transmission lines. The frameworks include a detailed depiction of different thermal power plants (e.g., efficiency, emission factor, year of installation, lifetime and planned decommissioning), intermittent and other renewable energy technologies, and storage. The model frameworks decide on dispatch and investments by using exogenous assumptions on the development of electricity demand, fuel prices, and techno-economical characteristics of the considered technologies. Within those characteristics, technological progress is depicted via vintage classes based on the commissioning year of the technologies.

\begin{table}[ht!]
\centering
\caption{Considered model frameworks.}
\footnotesize
\begin{tabular}{p{1.9cm} p{4.7cm} p{1.9cm} p{1.2cm} p{1.6cm}}
 \\
    Name & Institution & Type & Software & References \\\toprule
     DIMENSION & Institute of Energy Economics at the University of Cologne & Optimization & GAMS, CPLEX & \cite{Richter2011}  \\\midrule
     EUREGEN & ifo Institute for Economic Research & Optimization & GAMS, CPLEX & \cite{Weissbart19} \\\midrule
     E2M2 & Institute of Energy Economics and Rational Energy Use & Optimization & GAMS, CPLEX & \cite{Sun2013E2M2} \\\midrule
     urbs & Chair of Renewable and Sustainable Energy Systems & Optimization & Pyomo, Gurobi & \cite{urbs2018}
     \\\midrule
     HECTOR & Institute for Future Energy Consumer Needs and Behavior & Simulation & Vensim &  
     \cite{sheykhha2019hector}\\\bottomrule
\label{tab:frameworks}
\end{tabular}
\end{table}

\subsection{Data harmonization}
\label{subsec:data} 

A database developed within the 4NEMO project, which contains all essential data for modeling different futures of an interconnected Europe, i.e., scenario assumptions and technology data, is used (see \ref{App:Assumptions} for an overview of the most relevant numerical assumptions).

\paragraph{Scenario assumptions} The assumptions cover fuel and carbon prices, electricity demand, and net-transfer capacity (NTC) expansion. One out of the four scenarios developed within the 4NEMO project is used in this paper~\cite{Mier2020Costs}. The chosen scenario, \emph{Stagnation of the EU}, describes a collaborative future to reach ambitious climate targets. The development of emissions stringency and grid interconnections is retrieved from expert interviews using the cross-impact balance method \cite{weimer2006cross}. Fuel prices and electricity demand were quantified by using the computational general equilibrium (CGE) model PACE \cite{Boehringer04}. 
Despite improvements in energy efficiency, the electricity demand roughly doubles (+105\%) until 2050 due to direct and indirect electrification of transport, services, manufacturing, and energy intensive industries (see Table~\ref{tab:demand}). This increase is in the higher end of the range spanned by similar studies (+66\% in \cite{EC2020}, +80\%--150\% in the ambitious scenarios of \cite{EC2018}). Nevertheless, this scenario is used because the demand increase pushes the models to exploit more renewable resources and thus triggers differences in results which are interesting for the model comparison.

This scenario assumes that carbon prices (within the EU ETS) rise to 132 EUR/t\textsubscript{CO\textsubscript{2}} in 2050 (see Table~\ref{tab:carbonpr}) and that interconnectivity remains at the 2030 level in accordance to the 10-year development plan. This paper further considers the following variations:
\begin{itemize}
    \item \emph{lower CO\textsubscript{2} price}: Carbon price of 44 EUR/t\textsubscript{CO\textsubscript{2}} in 2050
    \item \emph{higher CO\textsubscript{2} price}: Carbon price of 176 EUR/t\textsubscript{CO\textsubscript{2}} in 2050
    \item \emph{increased NTC}: Interconnectivity targets beyond 2030 increase to 25\%
\end{itemize}

\paragraph{Technology data} This data covers information about existing infrastructure (generation, storage, and transmission capacities) as well as their projected expansion. 

\noindent The database provides assumptions about past (1960 to 2010), current (2015), and future (2020 to 2050) techno-economic parameters. It includes ten different thermal plants (lignite, coal, coal with CCS, nuclear, solid bioenergy, solid bioenergy with CCS, open-cycle gas turbines, combined-cycle gas turbines, steam turbines with natural gas, geothermal). The technologies differ with regards to their key technology-specific characteristics (efficiency, reliability\footnote{Reliability is constantly available capacity to reflect planned maintenance intervals and unplanned outages of power plants.}, capacity credit\footnote{Capacity credits are used to reflect the regulatory constraints of resource adequacy (firm capacity).}, CO\textsubscript{2} emission factor, CO\textsubscript{2} capturing factor for CCS technologies, commissioning year, lifetime, depreciation period, variable and fixed operations and maintenance cost, investment cost). Bioenergy and CCS technologies are restricted by country-specific bioenergy or carbon-storage potentials, respectively. In line with current political announcements, investments into nuclear power are restricted to Eastern Europe, France, Scandinavian countries, and the United Kingdom.

\noindent The database further contains eight different intermittent renewable technologies (hydro power plants, three different onshore wind and three different offshore technologies\footnote{The different wind technologies represent different hub heights of wind turbines, i.e., \SI{80}{\meter}, \SI{100}{\meter}, and \SI{120}{\meter}.}, solar PV) with potentials and hourly availabilities (i.e. capacity factors) based on \citet{Siala2020}. The hourly capacity factors are deterministic. Wind and solar technologies are differentiated into three resource quality classes -- high, medium, and low -- which reflect the geographic diversity of each region. The time series of each region were generated for multiple representative sites with predefined quality levels, then weighted so that the wind and solar generation are calibrated to historical data. Curtailment is endogenously determined and eventually reduces the yearly capacity factors of intermittent renewable technologies, which are limited by the harmonized availability factors for each technology, site, and resource quality class. This means that the models may not use all the renewable energy that is available in every time step, if its integration leads to higher costs, for example through the construction of transmission lines and storage. Finally, transmission capacity is approximated by using net-trade capacities.

\subsection{Major differences}
\label{subsec:majordiff}

The model frameworks are compared with regard to four major differences of power market models: (1) model type, (2) planning horizon, (3) temporal resolution, and (4) spatial resolution. 

\paragraph{Model type} There is a differentiation between optimization and simulation frameworks. Optimization frameworks decide on dispatch and investment by minimizing overall system costs (variable, fixed operation and maintenance, as well as investment cost). The considered simulation framework (HECTOR) determines dispatch based on variable costs, but uses a heuristic for investment decisions.

\noindent The considered optimization model frameworks (DIMENSION, EUREGEN, E2M2 and urbs) rely on the fundamental assumptions of perfectly competitive markets, symmetric firms with homogeneous information and cost structures, and no transaction cost. Representative firms seek to maximize their profits, which, under the above mentioned fundamental assumptions, is equivalent to a minimization of total system cost. Consequently, optimization models represent idealized depictions of real power markets. 

\noindent Simulation models allow to break the fundamental assumptions described above. For example, HECTOR abstracts from perfect competition, assuming that firms might skim off scarcity rents in hours where available capacity is scarce. In doing so, HECTOR tries to depict peak pricing in peak-load hours more realistically. Deviating from idealized markets, however, requires HECTOR to solve the underlying problem iteratively.

\paragraph{Planning horizon} The planning horizon reflects the foresight in investment behavior. Intertemporal models assume perfect foresight and thus can avoid path dependencies and lock-in effects by calculating dispatch and investment decisions considering the entire time horizon. Myopic models in turn do not avoid eventual path dependencies and lock-in effects. Such models divide the optimization problem into sub-problems, which are solved consecutively. Under myopia, dispatch and investment decisions in a certain period are made without the knowledge of future investment periods. The results obtained from the optimization of the respectively prior periods are taken as input to optimize the current period.

\paragraph{Temporal resolution} This feature defines how many time steps are used to reflect the fluctuations within a year. In general, two approaches can be differentiated: Fully depicting a year using 8760 hourly time steps, or using representative hours which are chosen heuristically or through clustering. A full depiction preserves all the information about the fluctuations of the load and renewable generation profiles. Representative time steps reflect the most significant patterns and reduce model complexity significantly. However, representative hours may struggle in accurately modeling storage behavior (\citet{nahmacher16}).

\noindent In this paper, a full year depiction (8760 hours) is compared to the use of 16 representative days (384 hours). Due to the computational complexity, intertemporal models were only able to perform the runs with 384 time steps. Simplified time series are scaled so that the annual demand and full-load hours of intermittent technologies match those of full temporal resolution. 

\paragraph{Spatial resolution} This feature describes the number and shape of the model regions. It is normally a flexible parameter that varies according to the research question. The most intuitive spatial resolution for power market models equals the spatial configuration of the depicted wholesale power markets. For most European countries, market configuration is congruent with national borders, while a few countries are split into several markets (e.g. Norway, Sweden).
 
\noindent A country-wise depiction (28 regions) is compared to a setting where the countries are aggregated into 12 mega-regions. Using the ISO 3166-1 alpha-2 codes: DE, FR, and IT are modeled as single countries. DK, FI, NO and SE are grouped into one region. Other regions are AT-CH, BE-LU-NL, BG-EL-RO, CZ-PO-SK, EE-LT-LV, ES-PT, HR-HU-SI, and IE-UK. 

\subsection{Deactivated and non-modeled features}
\label{subsec:deactfeatures}

First, several framework-specific features are deactivated for the sake of data harmonization. The focus is on wholesale power markets, and thus there is no depiction of other energy-related markets (e.g., balancing power or CO\textsubscript{2} markets) or other sectors (e.g., heat or transport). Framework-specific selection and weighting algorithms are deactivated since harmonized time series for electricity demand and intermittent renewable generation) with either 384 or 8760 time steps are used. 

Next, there is no depiction of direct subsidies (feed-in tariffs, market premiums, capacity payments) and indirect subsidies (renewable energy share targets, priority dispatch). In principle, all considered models are able to model different forms of subsidies although those features are rarely activated.
This assumption is particular valid for the European power market because the major climate policy that drives system transformation is the CO\textsubscript{2} price from the EU Emissions Trading System (EU ETS). 
The considered CO\textsubscript{2} price stems from a CGE calibration that seeks to model all policies projected by the European Joint Research Centre, that is, the price covers more than just the quantity target from the 
EU ETS but rather accounts for different implemented policies such as direct and indirect subsidies and taxes as well. Thus, the focus is on testing sensitivity of experiment outcomes for diverging CO\textsubscript{2} price paths without adding any overlapping, substituting, or even contrary policies such as direct and indirect subsidies, which often differ by country and might hamper the comparability of models.

Finally, technological constraints such as minimum dispatch, ramp-up restrictions, and unit commitment are not used. All considered models are principally able to depict minimum dispatch, but refrain from doing so because technology capacities are actually aggregated by vintages in each model (by the period of installation in five year steps). A minimum dispatch constraint is therefore often more restrictive for vintages that have sizes above the regular power plant size. 
\citet{gils2021modeling} show that such constraints have a minor impact and \citet{ramachandrana2021life} find that ramping restrictions do not matter for long-run capacity expansion. Regarding unit commitment, it cannot be depicted by any of the used models, at least not in their present specifications.
For such purpose, mixed-integer program unit commitment models are used \citep[e.g.,][]{bertsch2016flexibility}. 


\subsection{Remaining differences}
\label{subsec:remaindiff}

Some features are fundamental for the working of the respective model framework. Those features are kept active, although not all model frameworks were able to account for them.

\paragraph{Endogenous decommissioning}
Only the two model frameworks with intertemporal versions (DIMENSION and EUREGEN) allow to retire power plant, transmission, and storage capacities before the end of their technical lifetime.
\paragraph{Technology aggregation}
To ensure numerical feasibility, DIMENSION additionally reduces the number of vintages (for existing and newly built capacity) by merging every three vintages into one mega-vintage, for example the mega-vintage class 1960--1974 replaces 1960--1964, 1965--1969 and 1970--1974.
\paragraph{Storage}
Storage is deactivated in HECTOR. The framework is generally able to model it, but the way of doing so is incomparable with the way the other models do.
\paragraph{Negative emissions}
E2M2 is not capable of handling revenues from negative emission technologies. As a result, there is no benefit from investing in bioenergy with CCS, even though it is technically allowed.
\paragraph{Limits for renewable potentials}
The capacity expansion of solar and wind is limited by the total potential minus the existing capacities over all vintages. Since E2M2 cannot determine this variable endogenously, it sets exogenous upper bounds for each year that are constant over time.

\section{Model experiments}
\label{sec:experiments}
This section encompasses all the experiments used for the model comparison, starting by a benchmarking experiment followed by one experiment for each major model feature.

\subsection{Benchmarking experiment}
\label{subsec:bench}

\paragraph{Experiment setup}
Each model involved in this experiment uses a particular configuration of major features, which the modeler considers as basic setup. The models use the same harmonized input data for the scenario \emph{base}, modeling the European power market in 5-year steps between 2015 and 2050. Capacity expansion is only allowed after 2015.

\begin{table}[!htbp]
    \centering
    \caption{Models used in the benchmarking experiment.}
    \label{tab:benchmarking}
    \vspace{0.3cm}
    \footnotesize
    \begin{tabular}{c @{\hskip 15pt} c@{\hskip 5pt}c @{\hskip 15pt} c@{\hskip 5pt}c @{\hskip 15pt} c@{\hskip 5pt}c @{\hskip 15pt} c@{\hskip 5pt}c @{\hskip 15pt} c}
        Model variation & \rotatebox{90}{\underline{O}ptimization} & \rotatebox{90}{\underline{S}imulation} &
        \rotatebox{90}{\underline{I}ntertemporal} & \rotatebox{90}{\underline{M}yopic} &
        \rotatebox{90}{Full \underline{y}ear} & \rotatebox{90}{\underline{T}ypical steps} &  \rotatebox{90}{\underline{28} countries} &
        \rotatebox{90}{\underline{12} mega-regions} & Model framework\\
        \toprule
        OIT28-D & X & & X & & & X & X & & \underline{D}IMENSION \\[5pt]
        OIT12-R & X & & X & & & X & & X & EU\underline{R}EGEN \\[5pt]
        OMY12-E & X & & & X & X & & & X & \underline{E}2M2\\[5pt]
        OMY28-U & X & & & X & X & & X & & \underline{u}rbs \\[5pt]
        SMT12-H & & X & & X & & (X)\tablefootnote{HECTOR uses the same number of time steps, but not the exact same hours.} & & X & \underline{H}ECTOR\\[5pt]
        \bottomrule
    \end{tabular}
\end{table}

\begin{figure}[!htbp]
    \centering
    \includegraphics[width=\textwidth, trim={0 0 0 0.5cm}, clip]{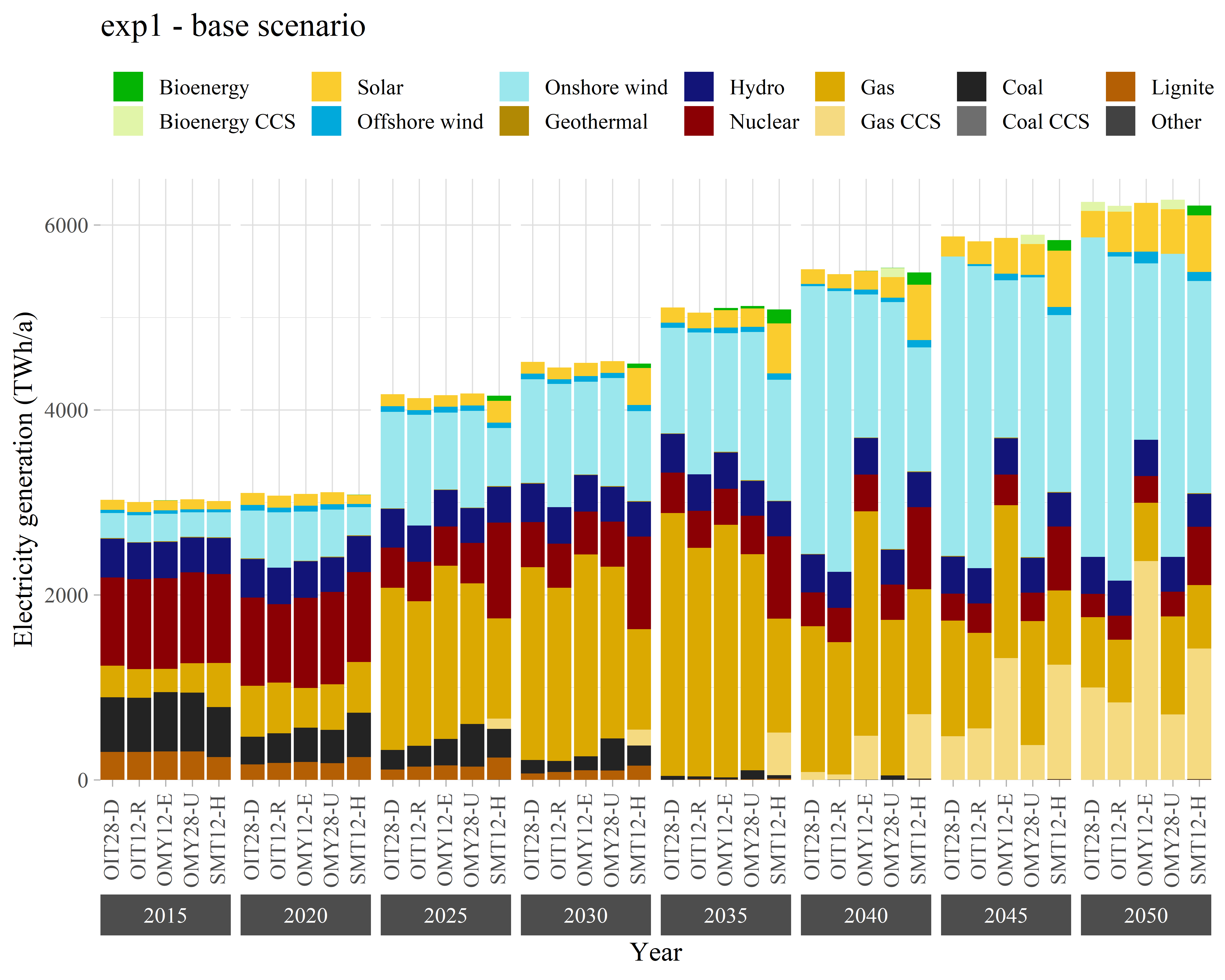}
    \caption{Electricity generation over the period 2015-2050 for all models involved in the benchmarking experiment.}
    \label{fig:benchmarking_mix}
\end{figure}

\paragraph{Results}
The comparison of the evolution of the electricity mix in Figure~\ref{fig:benchmarking_mix} shows that the results are not similar. Large differences start to appear in 2020, when endogenous expansion is allowed. The models disagree on when to phase out coal (OIT28-D: 2040, SMT12-H: beyond 2050), when to build gas CCS power plants (SMT12-H: 2025, OMY28-U: 2045), and on the share of onshore wind in 2050 (OMY12-E: 30.6\%, OIT28-D: 57.0\%), to name a few examples. The differences are also reflected in the total system costs and on the CO\textsubscript{2} emissions (see Supplementary materials~\cite{siala2020zenodo}).

\paragraph{Key message}
Since all models use the same harmonized input data, the differences must be due to the configuration settings of the model frameworks (major features) and/or due to the mathematical formulations of the specific model features. Therefore, the next experiments are designed to estimate the impact of the model features on the results.

\subsection{Impact of model type}
\label{subsec:type}

\paragraph{Experiment setup}
Apart from the feature ``model type'', all the other major features are harmonized. The compared models are myopic, they use 384 typical time steps, and they consist of 12 mega-regions. Table~\ref{tab:exp_type} provides an overview of the used models.

\begin{table}[!htbp]
    \centering
    \caption{Models used in the model type experiment.}
    \label{tab:exp_type}
    \vspace{0.3cm}
    \footnotesize
    \begin{tabular}{c @{\hskip 15pt} c@{\hskip 5pt}c @{\hskip 15pt} c@{\hskip 5pt}c @{\hskip 15pt} c@{\hskip 5pt}c @{\hskip 15pt} c@{\hskip 5pt}c @{\hskip 15pt} c}
        Model variation & \rotatebox{90}{\underline{O}ptimization} & \rotatebox{90}{\underline{S}imulation} &
        \rotatebox{90}{\underline{I}ntertemporal} & \rotatebox{90}{\underline{M}yopic} &
        \rotatebox{90}{Full \underline{y}ear} & \rotatebox{90}{\underline{T}ypical steps} &  \rotatebox{90}{\underline{28} countries} &
        \rotatebox{90}{\underline{12} mega-regions} & Model framework\\
        \toprule
        OMT12-D & X & & & X & & X & & X & \underline{D}IMENSION \\[5pt]
        OMT12-R & X & & & X & & X & & X & EU\underline{R}EGEN \\[5pt]
        OMT12-E & X & & & X & & X & & X & \underline{E}2M2\\[5pt]
        OMT12-U & X & & & X & & X & & X & \underline{u}rbs \\[5pt]
        SMT12-H & & X & & X & & (X)\tablefootnote{HECTOR uses the same number of time steps, but not the exact same hours.} & & X & \underline{H}ECTOR\\[5pt]
        \bottomrule
    \end{tabular}
\end{table}

\begin{figure}[!htbp]
    \centering
    \includegraphics[width=\textwidth, trim={0 0 0 0.5cm}, clip]{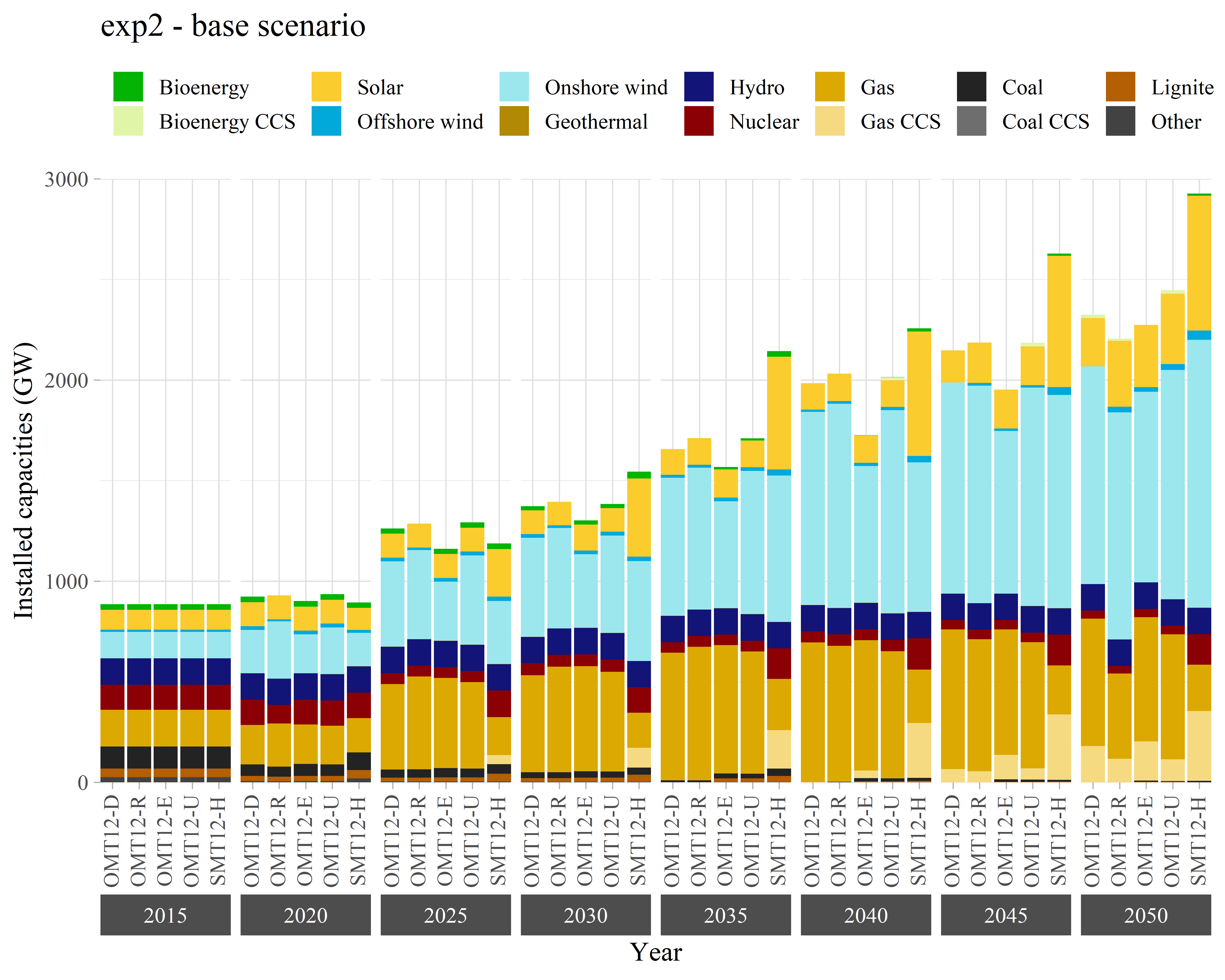}
    \includegraphics[width=\textwidth, trim={0 0 0 0.5cm}, clip]{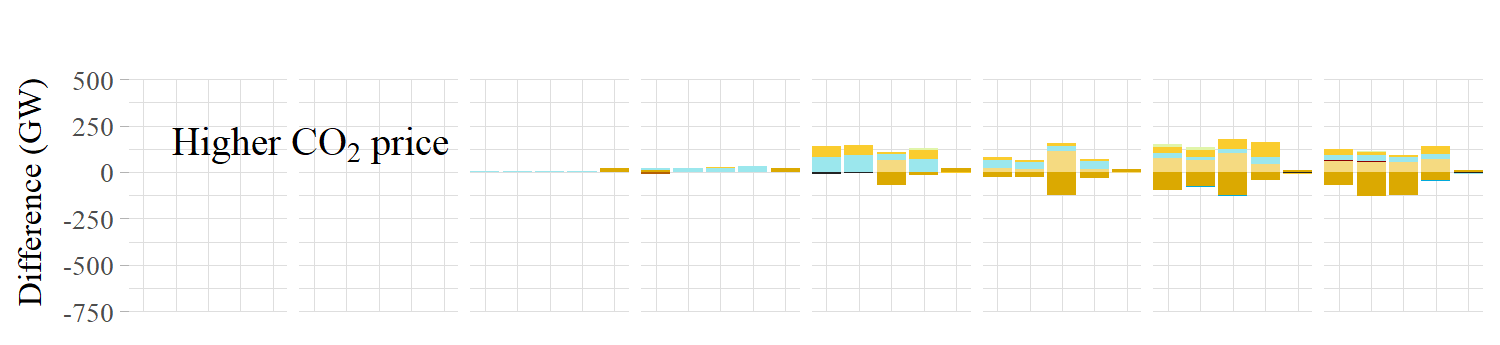}
    \includegraphics[width=\textwidth, trim={0 0 0 0.5cm}, clip]{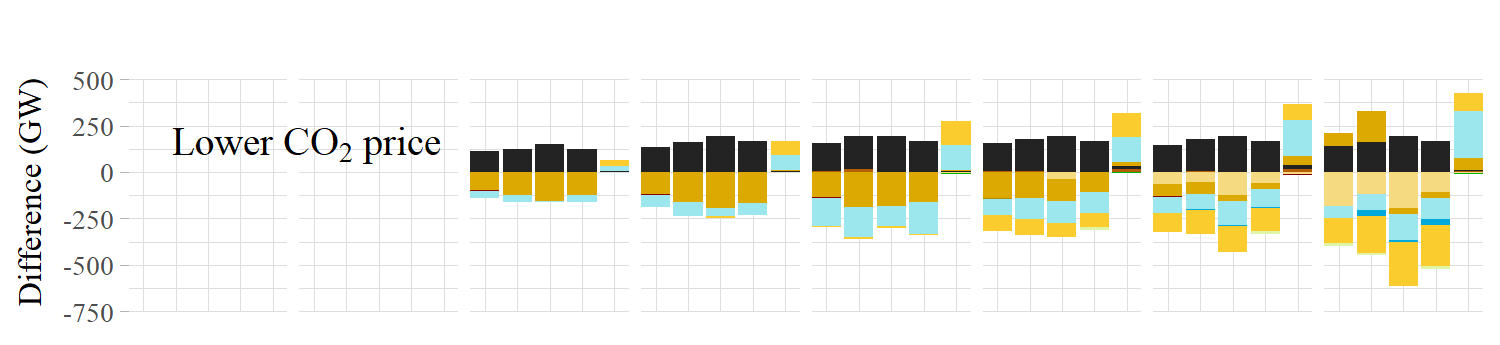}
    \caption{Total installed capacity in the period 2015-2050 for all models involved in the model type experiment (top) and differences due to higher and lower CO\textsubscript{2} prices (bottom).}
    \label{fig:type_instcap}
\end{figure}

\paragraph{Results}
The difference between the optimization models and the simulation model is most striking in the evolution of the installed capacities, which is displayed in Figure~\ref{fig:type_instcap}. Whereas myopic optimization models make investment decisions endogenously based on annualized total costs\footnote{Total costs include costs for investment, fix and variable O\&M, fuel, and CO\textsubscript{2} tax.}, the simulation model HECTOR uses a heuristic investment decision module for new capacities. This module expands all technologies with a positive net present value (NPV) and assigns investment shares to them based on (i) the total capacity of each region, (ii) the ratio of expected NPV to total investment costs, and (iii) on user-defined parameters. The calculation of the NPV requires additional assumptions on the expected full-load hours of each power plant type. This explains why the simulation model starts investing in gas power plants with CCS from 2025, or why it keeps investing in new nuclear capacities until 2050, even though these power plants will probably operate less often than expected due to the high share of renewable generation. The non-optimal allocation of new capacities leads to higher total system costs, which exceed the average costs of optimization models by 50\% in 2050.

\noindent The results of the optimization models are similar, yet not identical. In particular, the non-modeling of negative CO\textsubscript{2} emissions in E2M2 justifies the absence of bioenergy CCS in 2050. Also, the exogenous constraint on the yearly increase of renewable capacities in the same model might explain why its results show a lower capacity of onshore wind by 2050.

\paragraph{Sensitivity to other scenarios}
If the CO\textsubscript{2} price increases faster, the optimization models build negative emissions technologies (bioenergy and gas CCS) as soon as 2035, and install more onshore wind instead of conventional gas power plants. If the price increase is moderate, no negative emissions technologies are built, coal is used until 2050, and less solar and onshore wind power plants are built. The simulation model is barely sensitive to the increase of CO\textsubscript{2} price, since its energy mix and the shares of new installed technologies remain unchanged. However, the absolute values of the installed capacities in the scenario with lower CO\textsubscript{2} price increase because of two mutually reinforcing effects. On the one hand, the installed capacities of conventional power plants (e.g., gas) increase because they are cheaper. On the other hand, a system with higher dispatchable generators allows to integrate more renewable energy capacities. This is due to the way market frictions and investment dynamics are depicted in HECTOR.

\paragraph{Key messages}
The type of the model is a core difference that has a large impact on the results. Whereas the simulation model in this experiment can replicate the operation of a diverse portfolio of technologies, its investment decisions are very different from optimization models. The system suggested by the simulation model includes innovative technologies (e.g., gas CCS) earlier, and relies on conventional technologies (e.g., coal) for a longer period. It is also less sensitive to CO\textsubscript{2} price variations. The differences are down to the fact that the simulation model relies on a large number of user-defined parameters for its investment module, rather than a difference in paradigms between optimization and simulation. \citet{capros2014european} observe a similar behavior regarding the integration of CCS and the deployment of nuclear and coal, and recommend performing sensitivity analyses to validate the assumptions of simulation models.

\noindent The diversity in modeling approaches reveals the synergy between them: simulation models could use optimization results to calibrate their parameters and reduce the number of runs, then deliver electricity prices and other outcomes that better reflect existing imperfections in power markets.

\clearpage
\subsection{Impact of planning horizon}
\label{subsec:horizon}

\paragraph{Experiment setup}
Apart from the feature ``planning horizon'', all the other major features are similar. Only models based on the frameworks DIMENSION and EUREGEN are compared, as only they provide the option to vary the planning horizon. The compared optimization models use 384 typical time steps and consist of 12 mega-regions. Table~\ref{tab:exp_horizon} provides an overview of the used models.

\begin{table}[!htbp]
    \centering
    \caption{Models used in the planning horizon experiment.}
    \label{tab:exp_horizon}
    \vspace{0.3cm}
    \footnotesize
    \begin{tabular}{c @{\hskip 15pt} c@{\hskip 5pt}c @{\hskip 15pt} c@{\hskip 5pt}c @{\hskip 15pt} c@{\hskip 5pt}c @{\hskip 15pt} c@{\hskip 5pt}c @{\hskip 15pt} c}
        Model variation & \rotatebox{90}{\underline{O}ptimization} & \rotatebox{90}{\underline{S}imulation} &
        \rotatebox{90}{\underline{I}ntertemporal} & \rotatebox{90}{\underline{M}yopic} &
        \rotatebox{90}{Full \underline{y}ear} & \rotatebox{90}{\underline{T}ypical steps} &  \rotatebox{90}{\underline{28} countries} &
        \rotatebox{90}{\underline{12} mega-regions} & Model framework\\
        \toprule
        OIT12-D & X & & X & & & X & & X & \underline{D}IMENSION \\[5pt]
        OMT12-D & X & & & X & & X & & X & \underline{D}IMENSION \\[5pt]
        OIT12-R & X & & X & & & X & & X & EU\underline{R}EGEN \\[5pt]
        OMT12-R & X & & & X & & X & & X & EU\underline{R}EGEN \\[5pt]
        \bottomrule
    \end{tabular}
\end{table}

\begin{figure}[!htbp]
    \centering
    \includegraphics[width=\textwidth, trim={0 0 0 0.5cm}, clip]{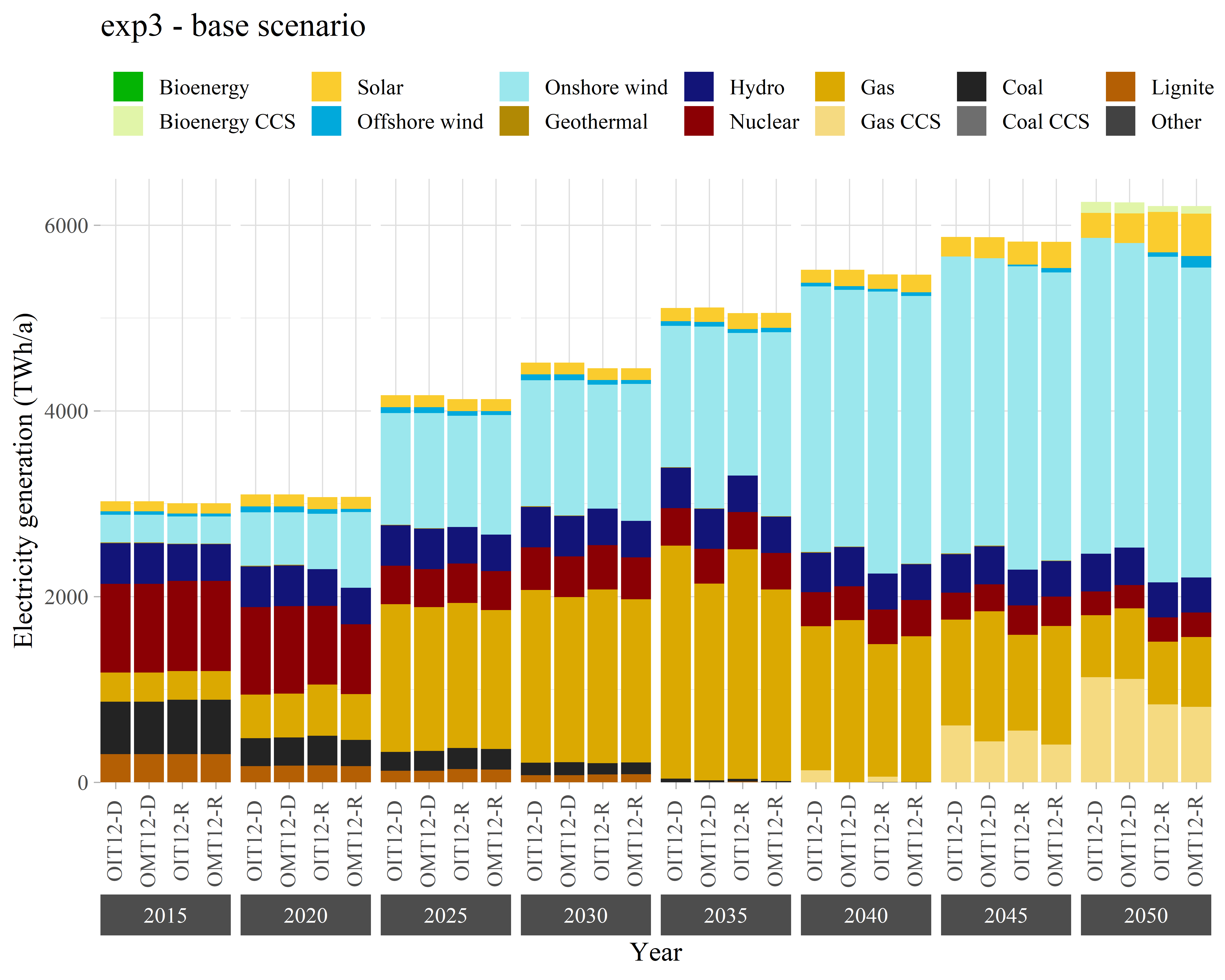}
    \includegraphics[width=\textwidth, trim={0 0 0 0.5cm}, clip]{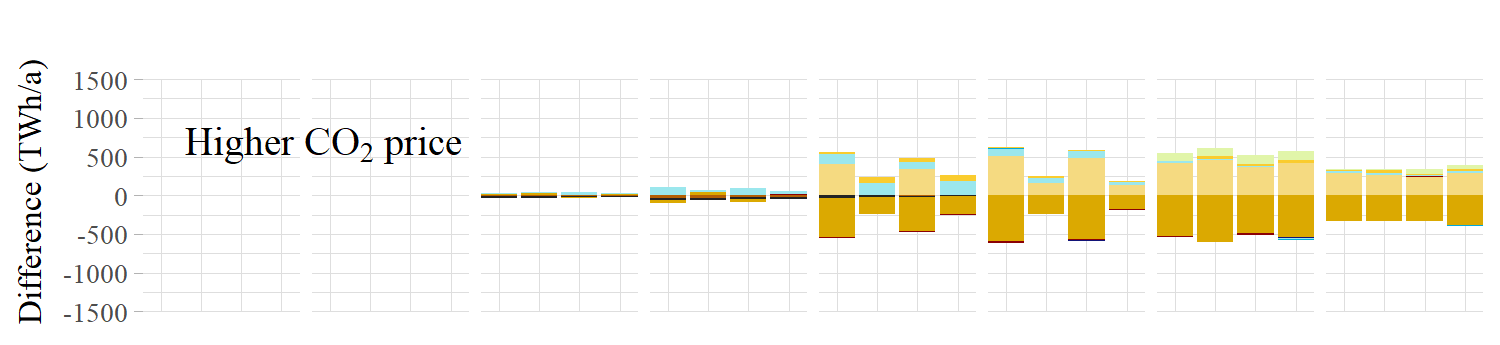}
    \includegraphics[width=\textwidth, trim={0 0 0 0.5cm}, clip]{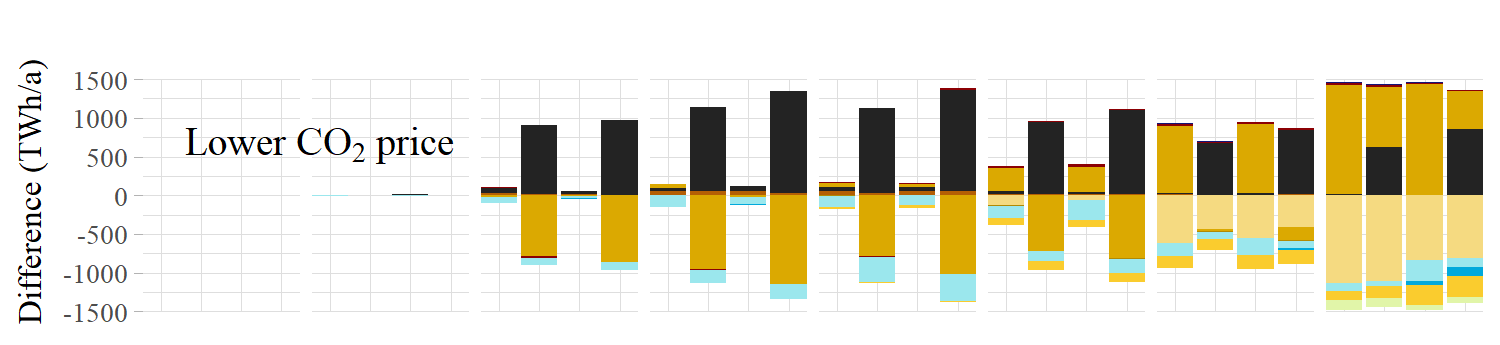}
    \caption{Electricity generation over 2015-2050 for the models involved in the planning horizon experiment (top) and differences due to higher and lower CO\textsubscript{2} prices (bottom).}
    \label{fig:horizon_mix}
\end{figure}

\paragraph{Results}
Figure~\ref{fig:horizon_mix} compares the evolution of the electricity generation mixes of the four models. Intramodel differences are first analyzed by observing the results of each pair of models from the same framework.

\noindent Both in 2015 and 2050, the differences between the models of the same framework are minor (the largest differences in onshore wind generation in 2050 are of 120~TWh/a for DIMENSION and 170~TWh/a for EUREGEN in 2050, so less than 2\% and 2.7\% of the generation, respectively). However, there are discrepancies for the years in between. For both frameworks, the myopic versions have a higher share of onshore wind between 2025 and 2035, which the intertemporal models replace with gas. From 2040 onward, the model OIT12-D builds gas CCS plants and uses more onshore wind and less conventional gas power plants than its myopic counterpart. OIT12-R also builds onshore wind at a large scale after 2040. This could be explained by the limits on the potential of onshore wind expansion. Since the intertemporal models have perfect foresight, they do not exhaust the wind potential until a new generation of more efficient wind turbines becomes available in 2040. This is at odds with other papers that show intertemporal models investing earlier in low-carbon technologies \citep{Heuberger17}, but the outcomes can plausibly be justified by the exogenous technological progress (which is actually only possible if there is high adoption of the technology in previous years) and the potential constraints for onshore wind.

\noindent The intertemporal models minimize the costs over the whole time horizon, whereas the myopic ones optimize the system design and operation for each period separately. Consequently, the system costs over the period 2015--2050 must be higher in the myopic models than in the intertemporal models (or at least equal). This is confirmed in this experiment, albeit the difference is marginal (DIMENSION: +1.0\%, EUREGEN: +1.6\%).

\noindent The reasons for the different behaviors in DIMENSION and EUREGEN could be traced back to aspects that have not been harmonized, such as technology vintage aggregation in DIMENSION or the exact implementation of endogenous decommissioning.

\paragraph{Sensitivity to other scenarios}
If the CO\textsubscript{2} price increases faster, the myopic models deliver solutions that are +1.5\% (OMT12-D) and +2.1\% (OMT12-R) more expensive than their intertemporal counterparts, while the energy mix does not vary considerably. While myopic and intertemporal models show similar behavior in the base as well as in the +CO\textsubscript{2} scenario, the investment horizon becomes crucial under lower carbon prices. If the CO\textsubscript{2} price increases slowly, the model OMT12-D uses coal power plants to cover 10--20\% of the energy demand from 2035 onward. In that scenario, the difference in system costs between OMT12-D and OIT12-D reaches +2.3\%, and +2.5\% for OMT12-R and OIT12-R. Under myopia, low carbon prices in the short-run lead to investments in new coal power plants. Models with perfect foresight anticipate that these investments will not pay off over the lifetime of the power plants since carbon prices will increase. Thus, intertemporal models rely on gas power plants instead.

\paragraph{Key messages}
Intertemporal and myopic models might agree on the optimal energy mix in the long term, but they disagree on the way to get there. Due to their perfect foresight, intertemporal models can take into consideration constraints that span over time (e.g., renewable potential constraints), or that change over time (e.g., CO\textsubscript{2} price), and make investment decisions accordingly. The lack of this capability in myopic models leads to slightly higher overall costs, but the gap widens for scenarios with lower CO\textsubscript{2} constraints. Assuming that policy makers consider similar long-term scenarios, they could compare the costs in both models and decide whether the difference justifies a regulatory intervention to control the wind expansion, for example.

\clearpage
\subsection{Impact of temporal resolution}
\label{subsec:temporal}

\paragraph{Experiment setup}
Apart from the feature ``temporal resolution'', all the other major features are similar. The compared optimization models are myopic and consist of 12 mega-regions. They either use 8760 time steps (\emph{full year}) or 384 (\emph{typical steps}). For the latter case, four consecutive days from different seasons were chosen. The models use the same time steps. Only the frameworks E2M2 and urbs are involved in this experiment because they have the computational infrastructure to run optimizations with 8760 time steps. Table~\ref{tab:exp_temporal} provides an overview of the used models.

\begin{table}[!htbp]
    \centering
    \caption{Models used in the temporal resolution experiment.}
    \label{tab:exp_temporal}
    \vspace{0.3cm}
    \footnotesize
    \begin{tabular}{c @{\hskip 15pt} c@{\hskip 5pt}c @{\hskip 15pt} c@{\hskip 5pt}c @{\hskip 15pt} c@{\hskip 5pt}c @{\hskip 15pt} c@{\hskip 5pt}c @{\hskip 15pt} c}
        Model variation & \rotatebox{90}{\underline{O}ptimization} & \rotatebox{90}{\underline{S}imulation} &
        \rotatebox{90}{\underline{I}ntertemporal} & \rotatebox{90}{\underline{M}yopic} &
        \rotatebox{90}{Full \underline{y}ear} & \rotatebox{90}{\underline{T}ypical steps} &  \rotatebox{90}{\underline{28} countries} &
        \rotatebox{90}{\underline{12} mega-regions} & Model framework\\
        \toprule
        OMT12-E & X & & & X & & X & & X & \underline{E}2M2 \\[5pt]
        OMY12-E & X & & & X & X & & & X & \underline{E}2M2 \\[5pt]
        OMT12-U & X & & & X & & X & & X & \underline{u}rbs \\[5pt]
        OMY12-U & X & & & X & X & & & X & \underline{u}rbs \\[5pt]
        \bottomrule
    \end{tabular}
\end{table}

\begin{figure}[!htbp]
    \centering
    \includegraphics[width=\textwidth, trim={0 0 0 0.5cm}, clip]{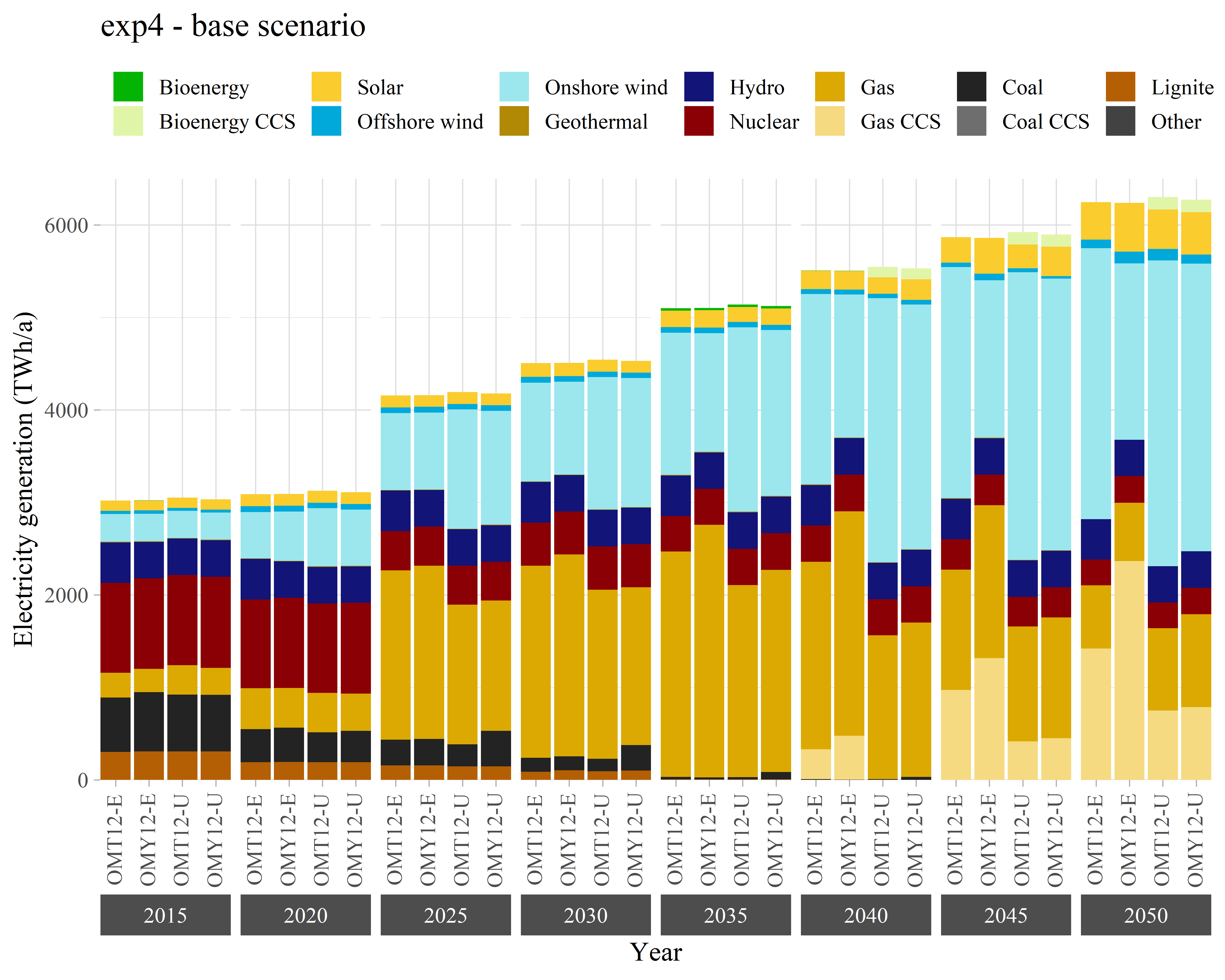}
    \includegraphics[width=0.99\textwidth, trim={0 0 0 0.5cm}, clip, right]{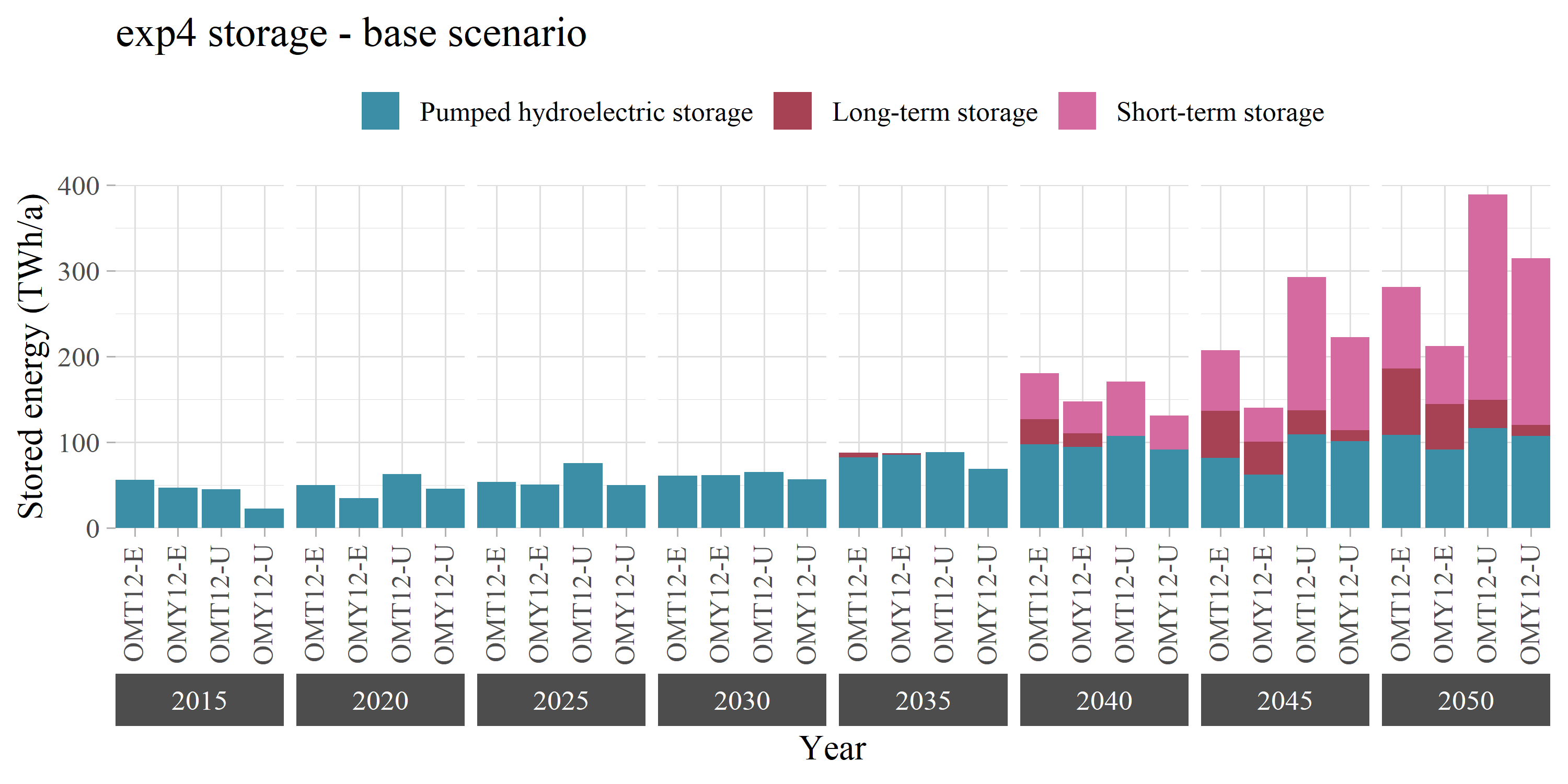}
    \caption{Electricity generation (top) and energy stored (bottom) over the period 2015-2050 for all models involved in the temporal resolution experiment.}
    \label{fig:temporal_mix}
\end{figure}

\paragraph{Results}
Analogously to the previous experiment, the intramodel differences are first analyzed by observing the results of each pair of models from the same framework. Figure~\ref{fig:temporal_mix} compares the evolution of the electricity generation mixes of the four models, and their usage of storage devices.

\noindent In both frameworks, it is observed that the models with fewer time steps rely more on onshore wind generation, which the 8760-time-step models replace in part with gas, gas CCS and solar generation. For instance, the share of onshore wind in electricity generation in 2050 is 47\% in OMT12-E and 30.6\& in OMY12-E, whereas the combined share of gas, gas CCS and solar shifts from 40,2\% to 56.6\%, respectively. The higher reliance on onshore wind occurs in conjunction with a higher usage of storage, in particular long-term storage. In fact, the share of long-term storage in OMT12-E and OMT12-U (27.8\% and 8.5\%, respectively) is greater than its share in OMY12-E and OMY12-U (25\% and 4.1\%, respectively). In absolute terms, most of the storage occurs in pumped hydroelectric storage up until 2040 (63.5\% and 69.5\% in 2040 in OMT12-U and OMY12-U, respectively), then in short-term storage devices (61.7\% and 61.6\% in 2050 in OMT12-U and OMY12-U, respectively).

\noindent The discrepancy between the models has two explanations. First and foremost, the reduction of the number of time steps necessarily impacts the shape of the time series, i.e., their hourly fluctuations. In this experiment, the time steps were selected heuristically, not through clustering and weighting of the clustered patterns. The time series were scaled and stretched to ensure that the total electricity demand and the full load hours of wind and solar generation are preserved. This process alters the shape of the time series, and affects each one independently of the others, so that the temporal correlation between weather and load events is not preserved. Clearly, this effect has favored onshore wind generation in this experiment.

\noindent Second, the way storage is modeled has an impact on the results. In all four models, the equations describing the storage state depend on its state in the previously modeled time step. In the absence of a mechanism to account for gaps between physically distant time steps in the models with 384 time steps, transitions between seasons are neglected. Consequently, the models inevitably overestimate the need for long-term, seasonal storage, which also favors the use of onshore wind.

\noindent One major difference between E2M2 and urbs models consists in the modeling of wind expansion. E2M2 splits the wind expansion potential into evenly distributed yearly portions, which remain unused in the first years and are not transferred to future years. It compensates for the lack of wind generation from 2040 onwards by using gas CCS with long-term storage more often than urbs. The fact that the model uses long-term storage to boost the full load hours of gas CCS seems counter-intuitive, since short-term storage is much cheaper per MW and is more efficient. However, the high energy-to-power ratio of long-term storage (720 MWh/MW to 4 MWh/MW) allows a much more cost-effective shifting of energy despite higher storage losses. The  reliance on long-term storage makes E2M2 models more sensitive to the change of the temporal resolution.

\paragraph{Key messages}
The impact of the temporal resolution is correlated with the quality of the complexity reduction technique. If the reduction of the time steps does not preserve the temporal correlation of the demand and renewable supply time series, variations in the share of renewable technologies might occur. The experiment echoes the recommendations of \citet{nahmacher16} regarding the impact of the choice of time steps. In case irregular time gaps are not accounted for in storage modeling, the variations might be amplified. 

\subsection{Impact of spatial resolution}
\label{subsec:spatial}

\paragraph{Experiment setup}
Apart from the feature ``spatial resolution'', all the other major features are the same. All the compared optimization models are myopic and use 384 typical time steps. They either use 28 single countries or 12 mega-regions. HECTOR is not involved in this experiment because of its inability to handle a large amount of regions. Table~\ref{tab:exp_spatial} provides an overview of the used models.

\begin{table}[!htbp]
    \centering
    \caption{Models used in the spatial resolution experiment.}
    \label{tab:exp_spatial}
    \vspace{0.3cm}
    \footnotesize
    \begin{tabular}{c @{\hskip 15pt} c@{\hskip 5pt}c @{\hskip 15pt} c@{\hskip 5pt}c @{\hskip 15pt} c@{\hskip 5pt}c @{\hskip 15pt} c@{\hskip 5pt}c @{\hskip 15pt} c}
        Model variation & \rotatebox{90}{\underline{O}ptimization} & \rotatebox{90}{\underline{S}imulation} &
        \rotatebox{90}{\underline{I}ntertemporal} & \rotatebox{90}{\underline{M}yopic} &
        \rotatebox{90}{Full \underline{y}ear} & \rotatebox{90}{\underline{T}ypical steps} &  \rotatebox{90}{\underline{28} countries} &
        \rotatebox{90}{\underline{12} mega-regions} & Model framework\\
        \toprule
        OMT28-D & X & & & X & & X & X & & \underline{D}IMENSION \\[5pt]
        OMT12-D & X & & & X & & X & & X & \underline{D}IMENSION \\[5pt]
        OMT28-R & X & & & X & & X & X & & EU\underline{R}EGEN \\[5pt]
        OMT12-R & X & & & X & & X & & X & EU\underline{R}EGEN \\[5pt]
        OMT28-E & X & & & X & & X & X & & \underline{E}2M2 \\[5pt]
        OMT12-E & X & & & X & & X & & X & \underline{E}2M2 \\[5pt]
        OMT28-U & X & & & X & & X & X & & \underline{u}rbs \\[5pt]
        OMT12-U & X & & & X & & X & & X & \underline{u}rbs \\[5pt]
        \bottomrule
    \end{tabular}
\end{table}

\begin{figure}[!htbp]
    \centering
    \includegraphics[width=\textwidth, trim={0 0 0 0.1cm}, clip]{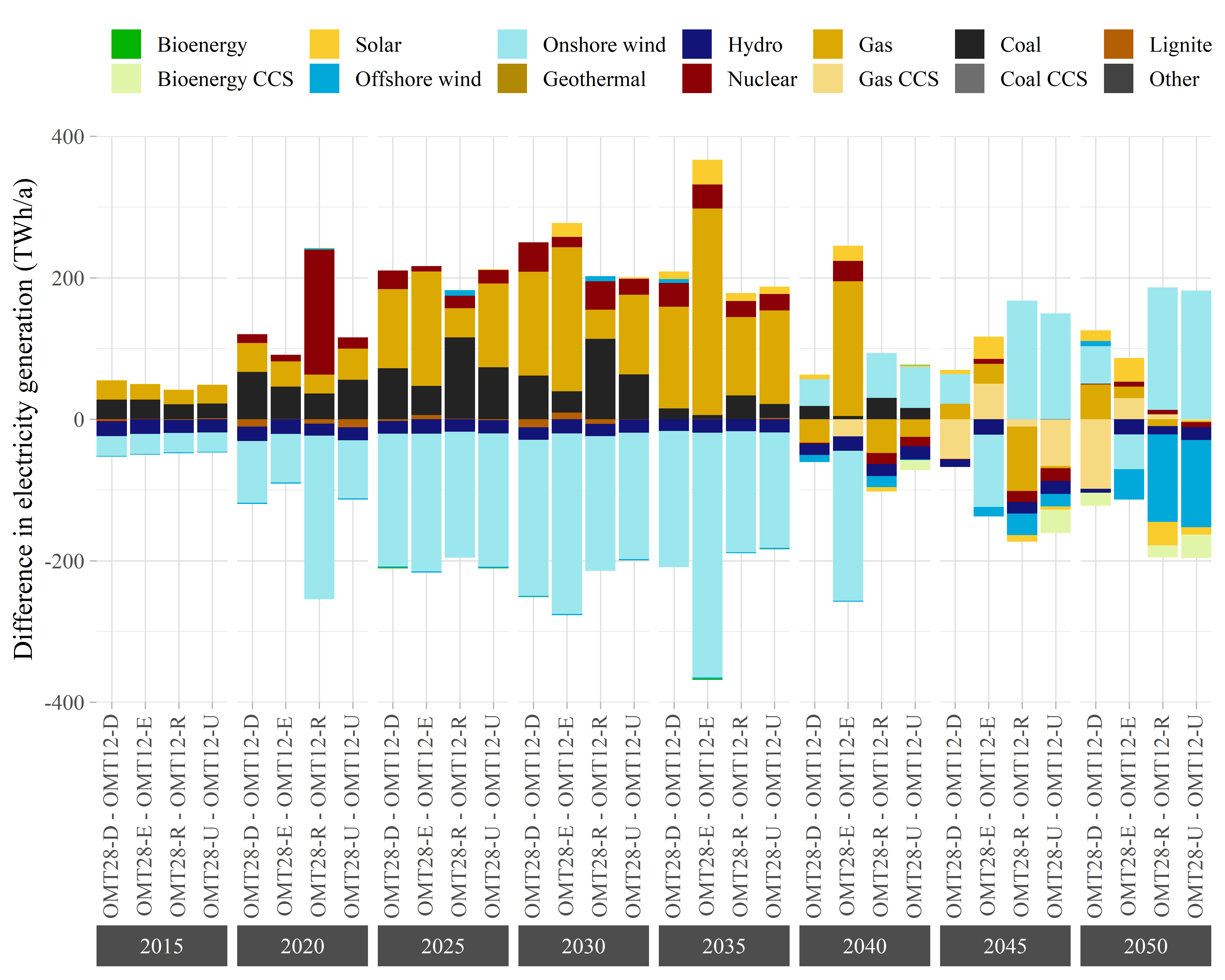}
    \includegraphics[width=\textwidth, trim={0 0 0 0.5cm}, clip]{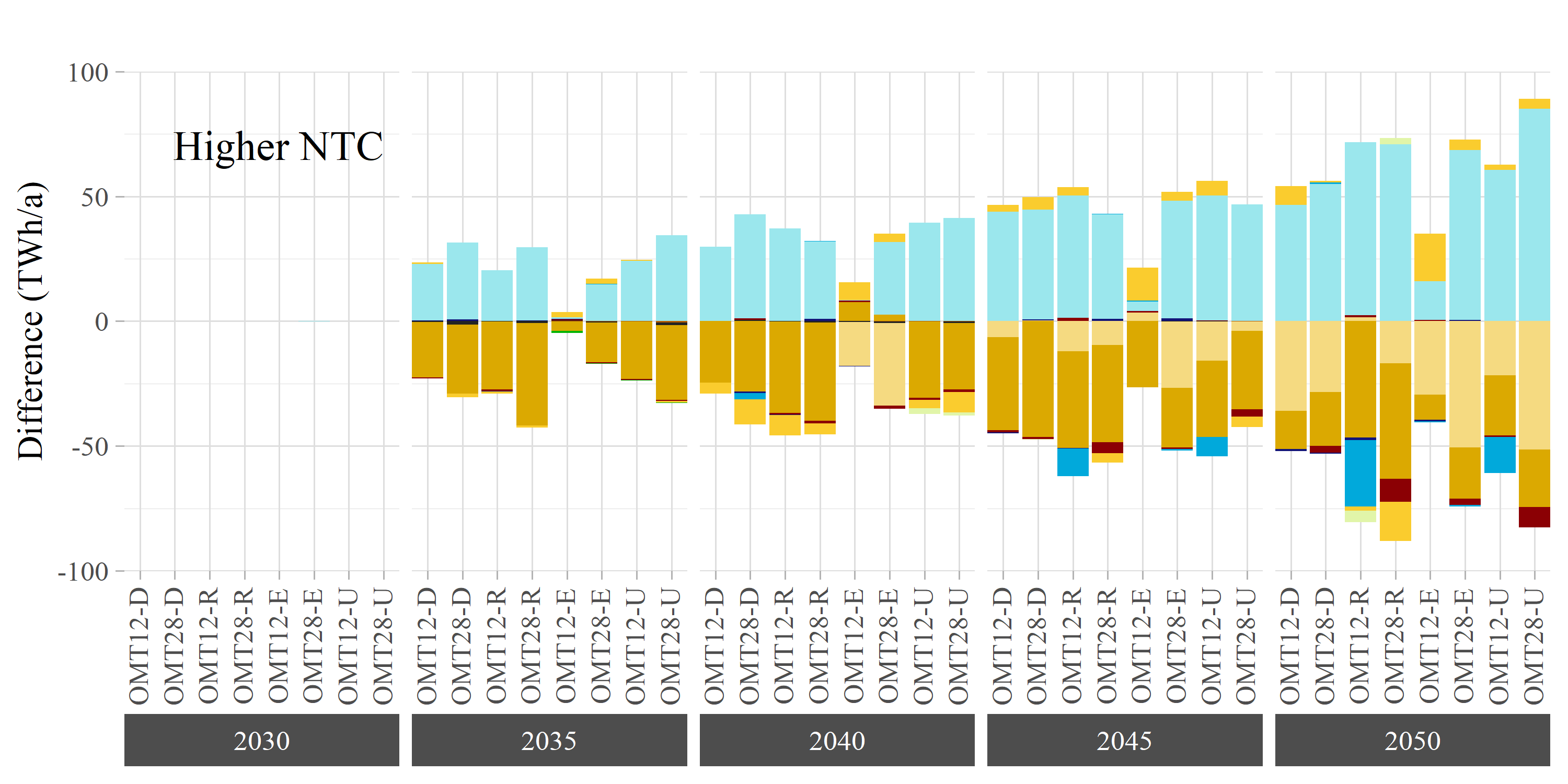}
    \caption{Difference in electricity generation between models with 28 and 12 regions over the period 2015-2050 (top) and difference due to increased net transfer capacities beyond 2030 (bottom) for all pairs of models involved in the spatial resolution experiment.}
    \label{fig:spatial_mix}
\end{figure}

\paragraph{Results}
Due to the large number of models involved in this experiment, the results regarding the energy mix were aggregated in Figure~\ref{fig:spatial_mix} for each pair of similar models. The difference between the two models for each framework is shown, meaning that positive values stand for a higher electricity generation using the depicted technologies in the models with 28 regions, while values below zero indicate that the generation in the models with 12 regions is higher. Up until 2035 (for E2M2) or 2040 (for the rest), the energy mix is characterized by an over-estimation of onshore wind generation in models with 12 regions. In models with a higher spatial resolution, this is replaced by gas, coal, nuclear, and solar. In the long term, models with 28 regions rely more on onshore wind and less on offshore wind, gas CCS, and bioenergy CCS. In absolute terms, the discrepancies could reach up to 370 TWh in 2035, which is about 7\% of the energy demand of that year.

\noindent Similarly to the previous experiment, the the spatial aggregation deals with the weather data and the load separately, so that the spatial and temporal correlation between weather and load events is not preserved, which could partly explain the discrepancies. However, unlike the reduction of the temporal complexity, the spatial reduction does not preserve the full load hours of renewable generation either. In fact, when grouping two or more regions, the distribution of the potentials shifts, and the locations of the representative sites (e.g. median, upper 10\%) also change. This results in completely different time series for the aggregated regions, both in shape (hourly fluctuations) and in sum (yearly full load hours). The weighting of the time series calibrates the wind and solar generation to historical data, but does not constrain future expansion. In most cases, this has led to an over-estimation of the wind potential in very large regions. Also, in the absence of transmission constraints between the countries making up each mega-region, the integration of wind is facilitated. This explains why the models with 12 regions tend to have more onshore and offshore wind. By 2040, the potential for good wind sites is already exhausted in most 12-region models. Meanwhile, the models with 28 regions profit from the existence of more efficient wind turbines to decarbonize faster, hence their higher onshore wind generation in the long term.

\paragraph{Sensitivity to other scenarios}
Increasing the net transfer capacities makes it easier to integrate renewable energy generation from solar and onshore wind (and nuclear power, in the case of OMT12-R). Since models with a higher spatial resolution have more constraints on inter-regional trade, they usually benefit more from higher transfer capacities, which ease such constraints. EUREGEN generates more onshore wind in the high resolution model, but not any extra nuclear energy, such that it benefits less from the increased NTCs overall.

\paragraph{Key messages}
The impact of the spatial resolution depends on the quality of the spatial aggregation of time series and transmission capacities. In this experiment, fewer regions led to more onshore and offshore generation. Easing the transmission constraints by increasing the net transfer capacities exogenously favored onshore wind, particularly in the high resolution models. The experiment echoes the recommendations of \citet{hess2018representing} regarding the modeling of the grid.

\section{Conclusion and discussion}
\label{sec:conclusion}

This paper assesses the impact of four major model features (model type, planning horizon, temporal resolution, spatial resolution) on the results of power market models. It uses five frameworks (DIMENSION, EUREGEN, E2M2, urbs, and HECTOR) for a qualitative and quantitative comparison. A high level of harmonization is ensured by using a joint input database and stripping the model frameworks from some of their specific features. A series of experiments is designed so that the impact of each major model feature is inspected separately. A combination of intramodel and intermodel comparisons is used in order to attribute deviations in results to the major feature analyzed.

Thanks to the high level of data and feature harmonization, clear conclusions on the behavior of the involved models could be derived from the experiments, in line with the initial objective. These conclusions are then translated into recommendations for modelers on how to minimize the discrepancies. The first finding relates to the choice of model type (optimization or simulation), which for the used models and setup leads to fundamental differences in the fleet of technologies. One possible recommendation is to use the differences synergetically through model coupling: Simulation models can build upon the results of an optimization to reduce the number of sensitivity iterations and answer specific questions that cannot be analyzed within an optimization framework. Second, in the experiment on the planning horizon (intertemporal or myopic), the choice of the appropriate configuration depends on the underlying assumptions about future developments of prices and technologies. If the scenario constraints already put the system on a decarbonization pathway, myopic models could deliver similar results to intertemporal ones, without lock-in effects. In such a case, it is recommended to use the myopic models, which are less computationally intensive. Third, the experiments on the temporal resolution (8760 or 384 time steps) and the spatial resolution (28 countries or 12 mega-regions) reveal how sensitive the models are to the input data. The choice of the aggregation method of the time steps and/or model regions has an impact on the model results. In order to improve their robustness, models need to take into account the behavior of storage in non-consecutive time periods and to reflect grid bottlenecks within the regions. Otherwise, they risk overestimating the ease of integrating renewable energy, particularly seasonal wind generation. If the future energy systems contain large amounts of fluctuating energy sources, it is recommended to use high temporal and spatial resolutions to reflect the system's behavior and interdependencies correctly.

Even though similarity is expected in experiments with a high level of harmonization, it triggers a discussion on whether there is a need for a high number of energy system models, particularly from the perspective of funding institutions. This paper provides two arguments in favor of model plurality. First, a variety of modeling approaches allows to evaluate the bandwidth of outcomes due to model uncertainty, as done by \citet{Gillingham2015}. For considering the bandwidth of power system modeling, including more than one simulation model or other model types is desirable. Second, while the results are similar within this paper, model specialties that could drive results apart from each other were not included.

This analysis comes with caveats. Only one simulation model is included in our sample and only limited conclusion can be drawn out of that. A larger sample of simulation models would allow for additional insights. The possibility of overlapping optimization as balance between myopic and intertemporal planning horizons is also neglected. Regarding the model complexity reduction techniques, the role of different levels of technology richness is not analyzed. The focus was not on the algorithms for the selection (and weighting) of typical time steps, nor for clustering regions. These aspects could be addressed in future work. The models were also not compared based on their CPU or RAM usage. Using the same computer architecture for such a comparison would provide useful insights from the software perspective. Despite all the limitations, this paper could achieve the objectives within its scope and sets a higher harmonization standard for future model comparisons.


\section*{Contributions and Acknowledgements}

\noindent This work was funded by the Federal Ministry for Economic Affairs and Energy of Germany (BMWi) within the project \textit{Research Network for the Development of New Methods in Energy System Modeling (4NEMO)} under grant number 0324008A. The authors would like to thank Wolfgang Habla, Kristina Govorukha, Philip Mayer and Paul Kunz for developing and providing the used scenarios. The contribution of Christoph Weissbart, Wolfgang Habla and Frieder Borggrefe in collecting information and building the input database is gratefully acknowledged.

\section*{CRediT author statement}

\noindent
\textbf{Kais Siala:} Conceptualization, Investigation, Data curation, Writing - Original Draft, Visualization;
\textbf{Mathias Mier:} Conceptualization, Methodology, Investigation, Data curation, Writing - Original Draft;
\textbf{Lukas Schmidt:} Conceptualization, Methodology, Investigation, Writing - Original Draft; 
\textbf{Laura Torralba-Díaz:} Methodology, Investigation, Writing - Original Draft;
\textbf{Siamak Sheykhha:} Investigation, Writing - Original Draft;
\textbf{Georgios Savvidis:} Methodology, Investigation, Visualization.

\clearpage
\appendix
\setcounter{table}{0}
\clearpage

\section{Numerical assumptions}
\label{App:Assumptions}

\begin{table}[ht]
    \centering
    \caption{Range of fixed operation and maintenance costs of generation technologies for all vintages.}
    \label{tab:tech_param}
    \vspace{0.3cm}
    \footnotesize
    \begin{tabular}{p{3cm} >{\centering\arraybackslash}p{3cm} >{\centering\arraybackslash}p{3.5cm}}
    
    Technologies	& Fixed O\&M costs [EUR\textsubscript{2015}/kW/a] \\[5pt]\toprule
Nuclear	 & 264 \\[5pt]
Lignite	 & 66 \\[5pt]
Coal	 & 55 -- 60	\\[5pt]
Coal CCS	 & 99	-- 109 \\[5pt]
Gas CCGT	 & 34	\\[5pt]
Gas ST	 & 34 \\[5pt]
Gas OCGT	 & 17 \\[5pt]
Gas CCS	 & 32 \\[5pt]
OilOther	 & 33 \\[5pt]
Bioenergy	 & 197 --	212	\\[5pt]
Bio CCS	 & 121 -- 130 \\[5pt]
Geothermal	 & 440 -- 495 \\[5pt]
PV	& 8 --	15 \\[5pt]
Wind Onshore	& 0 \\[5pt]
Wind Offshore	& 76 -- 131 \\[5pt]
Pumped Hydro	 & 5.3 \\[5pt]
Short-Term Storage	 & 6.6 -- 21.1 \\[5pt]
Long-Term Storage &	 30.4 \\[5pt]
\bottomrule
\end{tabular}
\end{table}

\begin{table}[ht]
    \centering
    \caption{Full-load efficiency [\%] of generation technologies for vintage classes based on their commissioning year.}
    \label{tab:eff}
    \vspace{0.3cm}
    \footnotesize
    \begin{tabular}{c c c c c c c c c c }
    Technologies &	1965 &	1970 &	 1975 &	1980 &	1985 &	1990 &	1995 &	2000 &	2005  \\[5pt] \toprule
Bioenergy &	- &	- &	- &	- &	- &	20.0 &	20.0 &	20.0 &	20.0  \\[5pt]
Bio CCS &	- &	- &	- &	- &	- &	- &	- &	- &	-  \\[5pt]
Coal &	33.0 &	33.0 &	34.2 &	35.5 &	36.2 &	37.0 &	38.5 &	40.0 &	41.5  \\[5pt]
Coal CCS &	- &	- &	- &	- &	- &	- &	- &	- &	-  \\[5pt]
Gas CCGT &	40.0 &	40.0 &	41.0 &	42.0 &	46.1 &	50.0 &	52.6 &	54.9 &	56.5 \\[5pt]
Gas OCGT &	23.0 &	23.0 &	25.0 &	27.0 &	29.5 &	32.1 &	34.5 &	37.0 &	38.3  \\[5pt]
Gas ST &	40.0 &	40.0 &	41.0 &	42.0 &	46.1 &	50.0 &	52.6 &	54.9 &	56.5 \\[5pt]
Gas CCS &	- &	- &	- &	- &	- &	- &	- &	- &	-  \\[5pt] 
Lignite &	35.0 &	35.0 &	36.0 &	37.0 &	37.5 &	38.0 &	40.5 &	42.9 &	44.4  \\[5pt]
Nuclear &	40.0 &	40.0 &	41.0 &	42.0 &	46.1 &	50.0 &	52.6 &	54.9 &	56.5 	 \\[5pt]
OilOther &	30.9 &	30.9 &	30.9 &	30.9 &	30.9 &	30.9 &	30.9 &	30.9 &	30.9 	 \\[5pt]
Pumped Hydro	& 79.2 & 79.2 & 79.2 & 79.2 & 79.2 & 79.2 & 79.2 & 79.2 & 79.2   \\[5pt]
Short-Term Storage	& - & - & - & - & - & - & - & - &  -  \\[5pt]
Long-Term Storage	& - & - & - & - & - & - & - & - &  -  \\[5pt] \toprule
 Technologies & 2010 &	2015 &	2020 &	2025 &	2030 &	2035 &	2040 &	2045 &	2050 \\[5pt] \toprule 
Bioenergy & 20.0 &	20.0 & 20.4 &	20.9 &	21.3 &	21.3 &	22.1 &	22.1 &	23.0 \\[5pt] 
Bio CCS & - &	20.0 &	20.0 &	20.0 &	20.0 &	20.0 &	20.0 &	20.0 &	20.0 \\[5pt]
Coal & 42.9 &	44.8 &	46.8 &	48.0 &	49.1 &	49.1 &	49.1 &	49.1 &	49.1 \\[5pt]
Coal CCS & - &	44.9 &	44.9 &	44.9 &	44.9 &	44.9 &	44.9 &	44.9 &	44.9 \\[5pt]
Gas CCGT & 58.1 &	59.2 &	60.1 &	61.1 &	62.1 &	62.1	 & 62.1 &	62.1 &	62.1 \\[5pt] 
Gas OCGT & 39.5 &	41.7 &	43.6 &	44.5 &	45.5 &	46.0 &	46.5 &	46.5 &	46.5 \\[5pt] 
Gas ST & 58.1 &	59.2 &	60.1 &	61.1 &	62.1 &	62.1 &	62.1 &	62.1 &	62.1 \\[5pt] 
Gas CCS & - &	59.1 &	59.1 &	59.1 &	59.1 &	59.1 &	59.1 &	59.1 &	59.1 \\[5pt] 
Lignite & 45.9 &	47.8 &	- &	- &	- &	- &	- &	- &	- \\[5pt]
Nuclear & 58.1 &	59.2 &	60.1 &	61.1 &	62.1 &	62.1 &	62.1 &	62.1 &	62.1 \\[5pt]
 OilOther & 30.9 &	30.9 &	- &	- &	- &	- &	- &	- &	- \\[5pt]
Pumped Hydro	& 79.2 & 79.2 & 79.2 & 79.2 & 79.2 & 79.2 & 79.2 & 79.2 & 79.2   \\[5pt]
Short-Term Storage	& - & 90.0 & 90.0 & 90.0 & 90.0 & 90.0 & 90.0 & 90.0 & 90.0  \\[5pt]
Long-Term Storage	& - & 36.0 & 36.0 & 36.0 & 36.0 & 36.0 & 36.0 & 36.0 & 36.0 \\[5pt] 
\bottomrule
\end{tabular}
\end{table}

\begin{table}[ht]
    \centering
    \caption{Development of carbon prices [EUR\textsubscript{2015}/t].}
    \label{tab:carbonpr}
    \vspace{0.3cm}
    \footnotesize
    \begin{tabular}{p{3cm} rrrrrrrr}
        Scenario & 2015 & 2020 & 2025 & 2030 & 2035 & 2040 & 2045 & 2050 \\[5pt]
        \toprule
        Base & 7.8 & 15.0 &	22.0 & 27.0 & 56.0 & 68.0 &	102.0 & 132.0   \\[5pt]
        Lower CO\textsubscript{2} price & 7.8 & 15.0 & 18.0 & 19.0 & 34.0 &	34.0 & 42.0 & 44.0   \\[5pt]
        Higher CO\textsubscript{2} price & 7.8 & 15.0 & 23.0 & 31.0 & 68.0 & 85.0 &	132.0 &	176.0  \\[5pt]
        \bottomrule
    \end{tabular}
\end{table}

\begin{table}[ht]
    \centering
    \caption{Development of investment costs [EUR\textsubscript{2015}/kW].}
    \label{tab:invcost}
    \vspace{0.3cm}
    \footnotesize
    \begin{tabular}{p{3cm} rrrrrrrr}
    Technology & 2015 & 2020 & 2025 & 2030 & 2035 & 2040 & 2045 & 2050 \\[5pt]
    \toprule
    Nuclear &	6600 &	6006 &	5346 &	5082 &	4818 &	4488 &	4488 &	4356 \\[5pt]
    Lignite &	1640 &	1640 &	1640 &	1640 &	1640 &	1640 &	1640 &	1640 \\[5pt]
    Coal &	1500 &	1500 &	1440 &	1410 &	1395 &	1380 &	1380 &	1365 \\[5pt]
    Coal CCS &	3415 &	3415 &	3278 &	3210 &	3176 &	3142 &	3142 &	3108 \\[5pt]
    Gas CCGT &	850 &	850 &	850 &	850 &	850 &	850 &	850 &	850 \\[5pt]
    Gas ST &	850 &	850 &	850 &	850 &	850 &	850 &	850 &	850 \\[5pt]
    Gas OCGT &	437 &	437 &	437 &	437 &	437 &	437 &	437 &	437 \\[5pt]
    Gas CCS &	1495 &	1495 &	1495 &	1495 &	1495 &	1495 &	1495 &	1495 \\[5pt]
    OilOther &	822 &	822 &	822 &	822 &	822 &	822 &	822 &	822 \\[5pt]
    Bioenergy &	4322 &	4236 &	4149 &	4149 &	4106 &	4063 &	4063 &	4020 \\[5pt]
    Bioenergy CCS &	4450 &	4361 &	4272 &	4272 &	4228 &	4183 &	4183 &	4139 \\[5pt]
    Geothermal &	12364 &	11993 &	11622 &	11498 &	11251 &	11127 &	11004 &	11004 \\[5pt]
    PV &	1300 &	1027 &	936 &	858 &	819 &	780 &	741 &	715 \\[5pt]
    Onshore wind &	1520 &	1397 &	1368 &	1339 &	1325 &	1310 &	1310 &	1296 \\[5pt]
    Offshore wind &	3600 &	3024 &	2700 &	2520 &	2376 &	2268 &	2160 &	2088 \\[5pt]
    Short-term storage &	1740 &	1740 &	1440 &	1120 & 1120 &	780	& 780 &	440
 \\[5pt]
    Long-term storage &	1520 &	1520 &	1520 &	1520 &	1520 &	1520	& 1520 &	1520
 \\[5pt]
    \bottomrule
    \end{tabular}
\end{table}

\begin{table}[ht]
    \centering
    \caption{Development of fuel prices [EUR\textsubscript{2015}/MWh\textsubscript{th}].}
    \label{tab:fuelpr}
    \vspace{0.3cm}
    \footnotesize
    \begin{tabular}{p{2.5cm} rrrrrrrr}
    Fuel & 2015 & 2020 & 2025 & 2030 & 2035 & 2040 & 2045 & 2050 \\[5pt]
    \toprule
    Uranium &	2.3 &	2.3 &	2.3 &	2.3 &	2.3 &	2.3 &	2.3 &	2.3 \\[5pt]
    Lignite &	7.0 &	7.0 &	7.0 &	7.0 &	7.0 &	7.0 &	7.0 &	7.0 \\[5pt]
    Coal &	8.3 &	8.2 &	8.1 &	7.9 &	7.8 &	7.7 &	7.6 &	7.5 \\[5pt]
    Natural gas &	20.7 &	20.3 &	20.0 &	19.7 &	19.3 &	19.0 &	18.8 &	18.5 \\[5pt]
    Oil &	40.3 &	40.8 &	41 &	41.2 &	41.4 &	41.8 &	42.3 &	42.6 \\[5pt]
    Bioenergy &	12.0 &	12.0 &	12.0 &	12.0 &	12.0 &	12.0 &	12.0 &	12.0 \\[5pt]
    \bottomrule
    \end{tabular}
\end{table}

\begin{table}[ht]
    \centering
    \caption{Development of electricity demand [TWh/a].}
    \label{tab:demand}
    \vspace{0.3cm}
    \footnotesize
    \begin{tabular}{p{1.5cm} rrrrrrrr}
    Country & 2015 & 2020 & 2025 & 2030 & 2035 & 2040 & 2045 & 2050\\[5pt]
    \toprule
AT &	63.5 &	64.3 &	78.3 &	91.1 &	137.1 &	146.9 &	155.5 &	163.0 \\[5pt]
BE &	83.1 &	82.2 &	96.6 &	107.3 &	131.5 &	158.0 &	181.6 &	196.2 \\[5pt]
BG &	29.6 &	30.1 &	34.7 &	36.0 &	37.3 &	40.4 &	42.7 &	44.4 \\[5pt]
CH &	58.2 &	60.8 &	67.2 &	70.5 &	116.3 &	126.0 &	136.3 &	147.5 \\[5pt]
CZ &	58.7 &	62.6 &	116.8 &	122.9 &	126.8 &	135.2 &	144.8 &	153.7 \\[5pt]
DE & 528.1 &	533.9 &	832.3 &	842.8 &	844.9 &	874.3 &	908.5 &	946.0 \\[5pt]
DK &	31.8 &	31.8 &	36.6 &	35.1 &	39.3 &	46.8 &	52.1 &	55.4 \\[5pt]
EE &	7.4 &	7.6 &	9.3 &	11.4 &	11.6 &	12.2 &	12.7 &	13.3 \\[5pt]
EL &	52.4 &	52.8 &	58.0 &	54.9 &	58.8 &	64.2 &	69.6 &	73.4 \\[5pt]
ES &	238.6 &	247.5 &	312.7 &	367.0 &	493.1 &	522.4 &	542.2 &	566.9 \\[5pt]
FI &	79.7 &	72.6 &	82.9 &	79.2 &	80.0 &	82.1 &	87.3 &	90.9 \\[5pt]
FR &	447.7 &	449.9 &	759.0 &	766.8 &	811.5 &	864.3 &	922.3 &	980.0 \\[5pt]
HR &	15.7 &	15.7 &	17.4 &	17.6 &	18.2 &	20.4 &	23.0 &	25.4 \\[5pt]
HU &	38.1 &	36.7 &	44.2 &	52.9 &	67.5 &	71.4 &	75.4 &	82.0 \\[5pt]
IE &  25.7 &	26.1 &	30.6 &	32.6 &	39.3 &	42.5 &	45.1 &	49.1 \\[5pt]
IT &	297.2 &	318.7 &	421.4 &	564.1 &	599.8 &	648.7 &	694.2 &	739.9 \\[5pt]
LT &	10.2 &	12.1 &	18.4 &	18.1 &	17.2 &	17.8 &	18.9 &	19.8 \\[5pt]
LU &	6.2 &	6.5 &	7.4 &	8.2 &	11.6 &	14.2 &	15.5 &	17.3 \\[5pt]
LV &	6.5 &	6.6 &	7.7 &	8.6 &	10.0 &	11.8 &	12.4 &	12.9 \\[5pt]
NL &	109.4 &	113.3 &	148.9 &	186.9 &	191.9 &	202.3 &	213.5 &	230.1 \\[5pt]
NO &	119.3 &	124.0 &	130.3 &	124.9 &	157.0 &	166.8 &	176.8 &	187.6 \\[5pt]
PL &	138.9 &	143.3 &	164.3 &	180.2 &	231.1 &	269.1 &	281.6 &	293.7 \\[5pt]
PT &	46.9 &	51.8 &	60.9 &	62.1 &	66.2 &	70.2 &	73.5 &	76.2 \\[5pt]
RO &	46.9 &	47.0 &	54.2 &	57.6 &	59.9 &	66.3 &	73.6 &	79.3 \\[5pt]
SE &	127.8 &	132.8 &	158.5 &	161.1 &	232.7 &	250.4 &	268.6 &	287.8 \\[5pt]
SI &	12.9 &	13.1 &	15.3 &	16.6 &	19.4 &	21.5 &	22.8 &	24.1 \\[5pt]
SK &	25.4 &	26.7 &	33.5 &	39.1 &	48.1 &	56.7 &	58.7 &	61.0 \\[5pt]
UK &	311.2 &	317.1 &	358.1 &	388.1 &	433.5 &	484.7 &	527.9 &	588.7 \\[5pt]
\bottomrule
    \end{tabular}
\end{table}

\setcounter{table}{0}
\clearpage

\section{Conducted runs}

The outputs for all model runs (data and plots) as well as the scripts for plotting are shared with open licenses in this repository for supplementary materials~\cite{siala2020zenodo}.

\begin{table}[ht!]
    \centering
    \caption{Overview of all the conducted model runs.}
    \label{tab:allruns}
    \vspace{0.3cm}
    \footnotesize
    \begin{tabular}{m{1.9cm}m{1.6cm} @{\hskip 15pt} c@{\hskip 5pt}c @{\hskip 15pt} c@{\hskip 5pt}c @{\hskip 15pt} c@{\hskip 5pt}c @{\hskip 15pt} c@{\hskip 5pt}c @{\hskip 15pt} c@{\hskip 5pt}c@{\hskip 5pt}c@{\hskip 5pt}c}
        Model framework & Model variation & \rotatebox{90}{\underline{O}ptimization} & \rotatebox{90}{\underline{S}imulation} &
        \rotatebox{90}{\underline{I}ntertemporal} & \rotatebox{90}{\underline{M}yopic} & \rotatebox{90}{Full \underline{y}ear} & \rotatebox{90}{\underline{T}ypical steps} &  \rotatebox{90}{\underline{28} countries} &
        \rotatebox{90}{\underline{12} mega-regions} & \rotatebox{90}{base} & \rotatebox{90}{higher CO\textsubscript{2} price} & \rotatebox{90}{lower CO\textsubscript{2} price} & \rotatebox{90}{increased NTC} \\[5pt]
        \toprule
        DIMENSION & OIT28-D & X & & X & & & X & X & & X & X & X & X \\[5pt]
        & OIT12-D & X & & X & & & X & & X & X & X & X & X \\[5pt]
        & OMT28-D & X & & & X & & X & X & & X & X & X & X \\[5pt]
        & OMT12-D & X & & & X & & X & & X & X & X & X & X \\[5pt]
        \midrule
        EUREGEN & OIT12-R & X & & X & & & X & & X & X & X & X & X \\[5pt]
        & OMT28-R & X & & & X & & X & X & & X & X & X & X \\[5pt]
        & OMT12-R & X & & & X & & X & & X & X & X & X & X \\[5pt]
        \midrule
        E2M2 & OMY12-E & X & & & X & X & & & X & X & & & \\[5pt]
        & OMT28-E & X & & & X & & X & X & & X & X & X & X \\[5pt]
        & OMT12-E & X & & & X & & X & & X & X & X & X & X \\[5pt]
        \midrule
        urbs & OMY28-U & X & & & X & X & & X & & X & & & \\[5pt]
        & OMY12-U & X & & & X & X & & & X & X & & & \\[5pt]
        & OMT28-U & X & & & X & & X & X & & X & X & X & X \\[5pt]
        & OMT12-U & X & & & X & & X & & X & X & X & X & X \\[5pt]
        \midrule
        HECTOR & SMT12-H & & X & & X & & (X)\tablefootnote{HECTOR uses the same number of time steps, but not the exact same hours.} & & X & X & X & X & X \\[5pt]
        \bottomrule
    \end{tabular}
\end{table}

\clearpage
\bibliographystyle{model1-num-names}
\bibliography{bibliography}

\end{document}